\documentclass[journal,9pt]{IEEEtran}

\usepackage{amsmath,amsfonts}
\usepackage{algorithmic}
\usepackage{algorithm}
\usepackage{array}
\usepackage{siunitx}

\usepackage[caption=false,font=normalsize,labelfont=sf,textfont=sf]{subfig}

\usepackage{textcomp}
\usepackage{stfloats}
\usepackage{url}
\usepackage{verbatim}
\usepackage{graphicx}
\usepackage{cite}
\usepackage{hyperref}
\usepackage{tikz}
\usepackage{pgfplots}
\usepackage{booktabs}

\usepackage{amsfonts}
\usepackage{amssymb}
\usepackage{xcolor}
\usepackage[english]{babel}
\usepackage[utf8]{inputenc}
\usepackage{listings}

\usetikzlibrary{spy}
\usetikzlibrary{positioning, arrows.meta, shapes.geometric}
\pgfplotsset{compat=1.17}

\begin{document}

\title{{A Microwave Imaging System for Soil Moisture  Estimation in  Subsurface Drip Irrigation }}

\author{\Large{Mohammad Ramezaninia  , Mohammad Zoofaghari}

\thanks{This paper was produced by the IEEE Publication Technology Group. They are in Piscataway, NJ.}
\thanks{Manuscript received August 1, 2023; revised September 21, 2023.}
\thanks{Michael Shell is with University of BCA. John Doe is with with University of ABC.}}

\maketitle

\begin{abstract}
The microwave imaging system(MIS) stands out among prominent imaging tools for capturing images of concealed obstacles. Leveraging its capability to penetrate through heterogeneous environments MIS has been widely used for subsurface imaging. Monitoring subsurface drip irrigation(SDI) as an efficient procedure in agricultural irrigation is essential to maintain the required moisture percentage for plant growth which is a novel MIS application. In this research, we implement a laboratory-scale MIS for SDI reflecting real-world conditions to evaluate leakage localization and quantification in a heterogeneous area. We extract a model to quantify the moisture content by exploiting an imaging approach that could be used in a scheduled SDI. We employ the subspace information of images formed by back projection and Born approximation algorithms for model parametrization and estimate the model parameters using a statistical curve fitting technique.  We then compare the performance of these imaging techniques in the presence of environmental clutter such as plant roots and pebbles. The proposed approach can well contribute to efficient mechanistic subsurface irrigation for which the local moisture around the root is obtained non-invasively and remotely with less than $20\%$ estimation error. 
\end{abstract}


\section{Introduction}\label{introduction}
\begin{figure}[htb]
	\centering
	\includegraphics[scale=1]{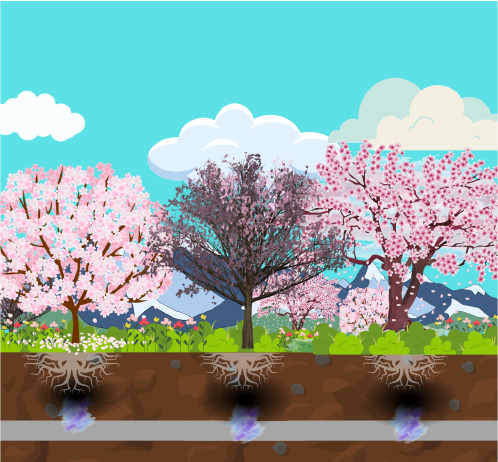}
	\caption{\small Schematic of subsurface drip irrigation (Generated by canva.com).}
	\label{ssee}
\end{figure}

\IEEEPARstart{S}{ubsurface} drip irrigation (SDI) is considered one of the latest techniques in mechanistic agriculture playing a crucial role in enhancing plant performance and efficiency, as well as reducing the global water consumption (Fig. \ref{ssee}). In this method, water is conveyed to the vicinity of plants roots  using regularly perforated pipelines buried shallowly beneath the ground. \cite{drip_irri}.
Real-time monitoring of moisture content around the plant roots is of interest in SDI. This would assist the farmer to adjust the specific wetness required for plant growth and avoid waste of water. In this regard, microwave imaging systems are a proper candidate especially for shallow underground imaging as they provide high-resolution images of the intended area. These images are post-processed to estimate the moisture content.

Satelite-based and Ground Penetrating Radar(GPR)-based approaches have been proposed for global and local Soil Moisture(SM) estimation respectively. 
Satellite-based approaches are widely introduced based on thermal, optical, and microwave remote sensing. The microwave sensing methods are stronger to penetrate the cloud and less sensitive to the land surface roughness. They usually exploit machine learning techniques considering various features of backscattered signals such as reflectivity and polarization.  These methods  more suit the global soil
moisture retrievals for meteorological and agricultural applications \cite{lei2020machine}. For example, authors in \cite{Bai}    introduce a methodology utilizing the first brightness Stokes parameter, which is the sum of vertical and horizontal polarization brightness temperature, as observed by the SM and Ocean Salinity (SMOS) satellite. The aim is to enhance SM retrieval accuracy, particularly in regions with complex topographic features.
On the other hand, GPR-based techniques for SM estimation have been proposed according to the soil permittivity variation versus moisture content. In \cite{9751623} a  deep artificial neural network (ANN) model is exploited to retrieve the SM based on the training data given by finite element simulations. Also, Anbazhagan et. al investigated the various equations to convert the dielectric coefficient into SM percentage \cite{dielec_wet}. They used a laboratory GPR on four different soil types to measure SM  considering a metal plate installed at a specific depth.
Most of these researches address the ean SM retrieval technique in the bare soil and for a wide area that is not applicable for SDI. 
Monitoring the SDI performance requires a local SM estimation in the presence of water pipelines, plant roots, and soil pebbles causing significant challenges in the imaging process.
For moisture detection in the vicinity of the pipeline, GPR-based approaches have been explored in the context of leakage detection as well.  \cite{mar16,marrr,var_gpr,mar19,mar18_sem2,mar22} 
For example, Amran et. al have employed a GPR operating at $800 \mathrm{MHz}$ to track the changes in the range-compressed images of leaking PVC pipes at various time intervals \cite{mar16}. Also,  a synthetic focusing algorithm is exploited in \cite{mar19} to investigate the image degradation along PVC pipes to locate the leakage area. Furthermore,  \cite{var_gpr} created a user-oriented 3D model of leakage detection utilizing GPR data processed by a second-order statistical filter examined for PVC pipe buried in a wooden soil box.  Many of these studies are limited to leakage detection but not the moisture estimation and quantification around the pipeline.

%

The local specification of SM could be explored within the realm of subsurface complex permittivity imaging \cite{Catapano}. Two main approaches have been considered: nonlinear inverse scattering methods and linear qualitative imaging techniques. Nonlinear inverse scattering is adept at determining the electrical parameters of scatterers, but its drawback lies in computational complexity and inefficient run time due to the iterative application of a forward scattering solver.  In \cite{Etminan} an inversion approach is designed to estimate root-zone SM by employing a novel optimization algorithm. The forward model employed in this paper to develop and evaluate the proposed inversion algorithm is the multilayered small perturbation method.
One example of this technique is the contrast source inversion method, extensively utilized in GPR models. Retrieving complex permittivity from GPR data poses a significant challenge due to the severely ill-posed nature of the problem. A novel imaging technique called reflection-damped plane-wave least-squares reverse time migration (RD-PLSRTM) aimed at improving scatter detection in migration images is introduced in \cite{Gao}. The inverse problem is resolved utilizing an iteratively reweighted least-squares algorithm. This method offers a versatile formulation that can be tailored to conventional PLSRTM and diffraction-based PLSRTM through the application of specific damping factors. Also, authors in \cite{Yamauchi} , propose an innovative approach to address this limitation by incorporating Ground Penetrating Synthetic Aperture Radar (GPSAR) into the contrast source inversion optimization scheme, which effectively reduces the number of unknowns. Yamauchi et. al in \cite{Yamauchi} exploited Green's function for the stratified heterogeneous media and tested their method using FDTD simulation data.

Qualitative subsurface imaging techniques (e.g. back projection, linear sampling method, diffraction tomography, etc.) create fast images of the intended area with low computational burden facilitating real-time object monitoring \cite{Ghaderi}. These techniques in contrast with nonlinear approaches cannot provide precise information about the object's permittivity. Moreover, in the case of a random heterogeneous medium like the moist soil in the vicinity of plant roots, it is impractical to establish a deterministic propagation model necessary for constructing the scattering operator in a qualitative inverse problem. In this regard, for example, the linear sampling method and time reversal MUSIC algorithm are utilized for image reconstruction of 2D objects in the stratified media with rough interfaces \cite{zoofaghari1,zoofaghari2}.  
The back projection has been studied as a qualitative microwave imaging in Ground Penetrating Synthetic Aperture Radar (GPSAR) as well. For example, \cite{Grathwohl} explores the impact of 3-D curved ground surface geometries on side-looking  GPSAR, to indicate that the neglect of tilt in BP is acceptable for imaging shallowly buried targets via unmanned aerial vehicles (UAVs).

Also, various microwave techniques have been employed for subsurface utility imaging. 
In \cite{mohan2016microwave} a microwave imaging system is introduced for moisture content estimation based on the transmission coefficient measured by the metamaterial-based sensor antenna located at different locations inside the soil. While this technique provides an accurate result, it involves several invasive operations that restrict the coverage area and prolong the imaging process.  Also, authors in \cite{Zhou} propose a novel approach for quantitative subsurface imaging using near-field scanning microwave microscopy with coaxial resonators, termed the learning-based (LB) method. The resolution enhancement by this method is confirmed by both numerical simulations and experimental data. This LB technique significantly reduces the runtime compared to conventional objective function methods as well. However, a key challenge lies in the generalization of the LB method to samples outside the training range, an issue not encountered by traditional approaches.

In this research, we utilized microwave imaging tools for capturing images of leakage areas 
and estimating the SM for SDI purposes. 
To this end, we extract a model for SM based on the subspace spectral coefficient of the clutter-removed reconstructed images in the predefined moist area. We employ two qualitative microwave imaging techniques to speed up the image formation process.  We relate the First Singular Value (FSV) of the image associated with the leakage region to the moisture content by the derived statistical fitting curves. The extracted model is verified considering the evaluation data and applied for real-time moisture monitoring in the vicinity of the plant root.  This leads to an inward moisture rate estimation that could be used in SDI scheduling for specified vegetation.

The general aspects of the proposed approach are outlined in Section \ref{proposed} and the image formation algorithms for the model extraction are discussed in Section \ref{image formation}. We detail the implemented setup and the scenarios of data acquisition in Section \ref{Senarios}. We also address various steps of model development, evaluation, and application considering the experimental data in this section. We conclude the paper in Section \ref{conclusion}.

\section{Proposed approach}\label{proposed}
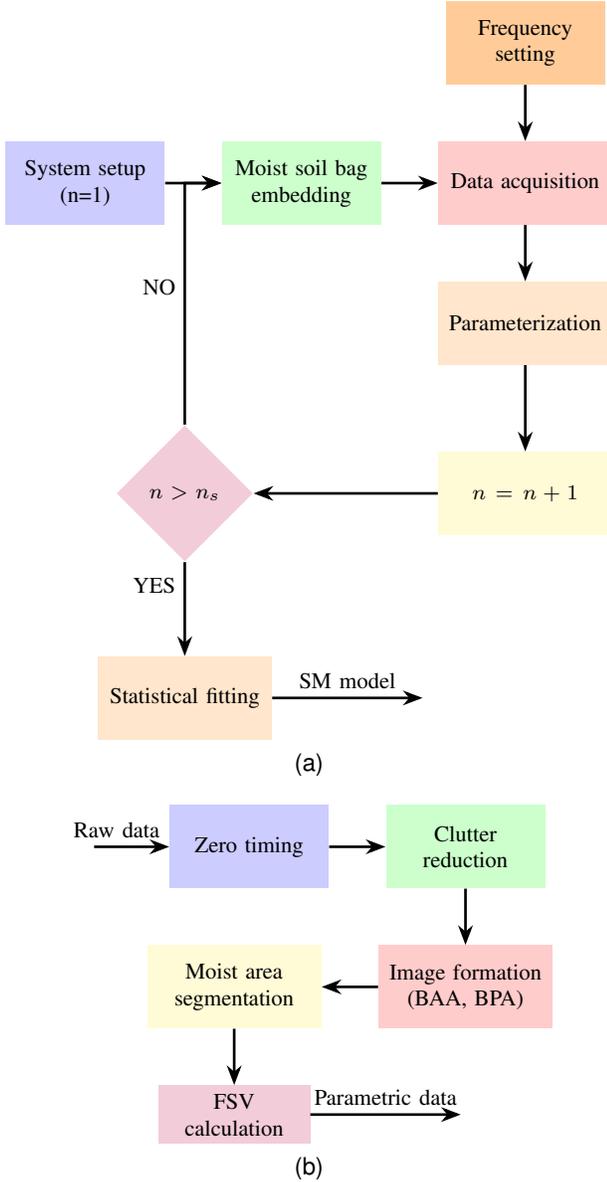
\begin{figure}[tb]
	\centering
	
	\subfloat[]{
		\begin{tikzpicture}[node distance=0.75cm, every node/.style={align=center, font=\small}, >=Stealth]
		\node (process1) [rectangle, minimum height=1.1cm, text width=1.9cm, fill=blue!20] {System setup\\(n=1)};
		\node (process2) [rectangle, minimum height=1.1cm, text width=1.9cm, right=of process1, fill=green!20] {Moist soil bag\\embedding};
		\node (process23) [rectangle, minimum height=1.1cm, text width=2.1cm, right=of process2, fill=red!20] {Data acquisition};
		\node (process233) [rectangle, minimum height=1.1cm, text width=1.9cm, above=of process23, fill=orange!40] {Frequency setting};
		\node (process3) [rectangle, minimum height=1.1cm, text width=2.1cm, below=of process23, fill=orange!20] {Parameterization};
		\node (process4) [rectangle, minimum height=1.1cm, text width=2.1cm, below=of process3, fill=yellow!20, yshift=-0.4cm] {$n = n + 1$};
		\node (process5) [diamond, minimum height=0.4cm, text width=1.2cm, left=of process4, fill=purple!20, xshift=-1.7cm] {$n>n_s$};
		\node (process6) [rectangle, minimum height=1.1cm, text width=2.1cm, below=of process5, fill=orange!20, yshift=-0.5cm] {Statistical fitting};
		
		\draw [->, line width=1pt] (process6.east) -- ++(2cm, 0) node[midway, above, draw=none] {SM model};
		\draw [->, line width=1pt] (process1) -- (process2);
		\draw [->, line width=1pt] (process2) -- (process23);
		\draw [->, line width=1pt] (process233) -- (process23);
		\draw [->, line width=1pt] (process23) -- (process3);
		\draw [->, line width=1pt] (process3) -- (process4);
		\draw [->, line width=1pt] (process4) -- (process5);
		\draw [->, line width=1pt, looseness=2, out=90] (process5) -- node[anchor=east, near start, draw=none] {YES} (process6);
		\draw [->, line width=1pt] (process5.north) -- ++(0,0) |- (process2.west) node[pos=0.25, anchor=south east, draw=none] {NO};
		\end{tikzpicture}
	}%
	\qquad
	\subfloat[]{
		\begin{tikzpicture}[node distance=0.75cm, every node/.style={align=center, font=\small}, >=Stealth]
		\node (process1) [rectangle, minimum height=1.1cm, text width=1.9cm, fill=blue!20] {Zero timing};
		\node (process2) [rectangle, minimum height=1.1cm, text width=1.9cm, right=of process1, fill=green!20] {Clutter reduction};
		\node (process3) [rectangle, minimum height=1.1cm, text width=2.1cm, below=of process2, fill=red!20] {Image formation\\(BAA, BPA)};
		\node (process4) [rectangle, minimum height=1.1cm, text width=2.1cm, left=of process3, fill=yellow!20] {Moist area\\segmentation};
		\node (process5) [rectangle, minimum height=0.4cm, text width=1.8cm, below=of process4, fill=purple!20] {FSV calculation};
		
		\draw [->, line width=1pt] ([xshift=-1cm]process1.west) -- (process1.west) node[midway, above, xshift=-0.2cm, draw=none] {Raw data};
		\draw [->, line width=1pt] (process5.east) -- ++(2cm, 0) node[midway, above, draw=none] {Parametric data};
		\draw [->, line width=1pt] (process1) -- (process2);
		\draw [->, line width=1pt] (process2) -- (process3);
		\draw [->, line width=1pt] (process3) -- (process4);
		\draw [->, line width=1pt] (process4) -- (process5);
		\end{tikzpicture}
	}%
	
	\caption{(a) Flowchart of proposed approach for SM model extraction using $n_s$ wet soil bags with known SM contents.  (b) Steps of SM model parametrization considering the  measured B-scan data.}
	\label{main_flowchart}
\end{figure}

\begin{figure}[htb]
\centering
\includegraphics[scale=0.23]{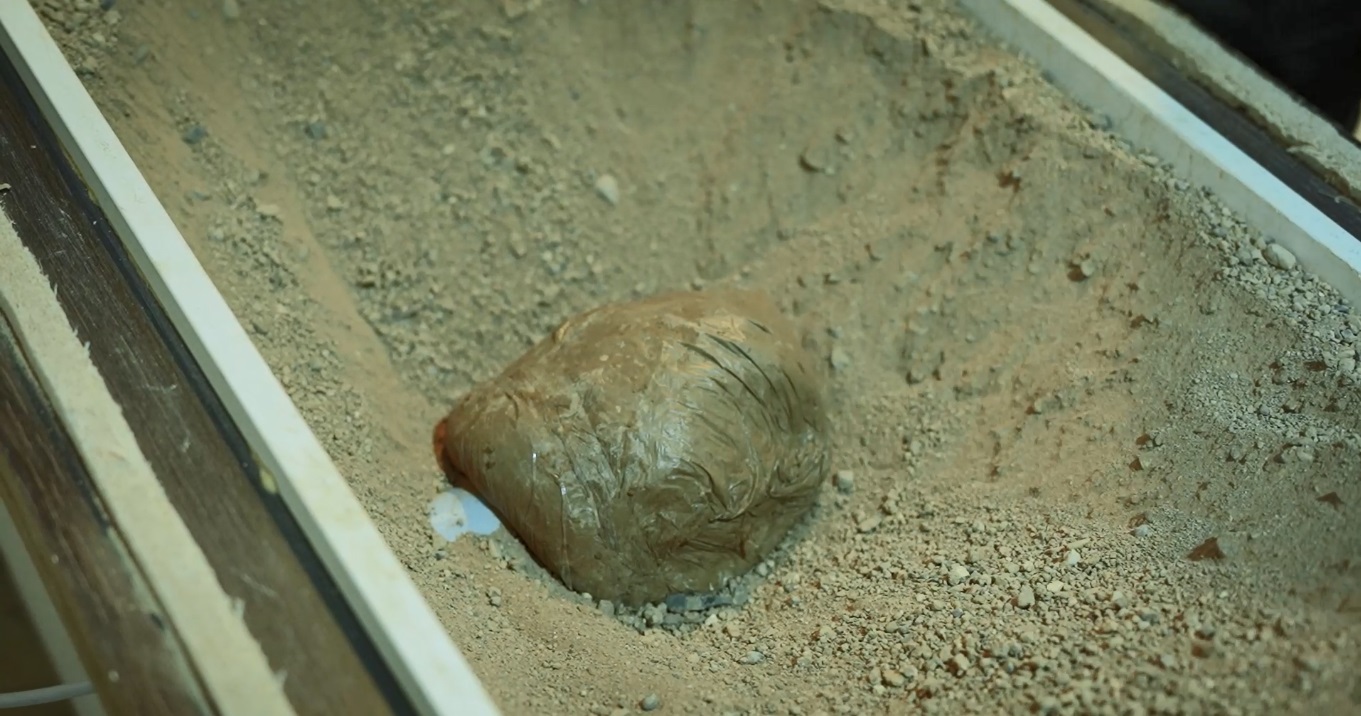}
\caption{\small Moist soil bag embedded over the PVC pipe.}
\label{one_pipe2}
\end{figure}

In the context of an SDI system, the greater the fluctuation
in SM over the pipe, the more pronounced
the deterioration of the pipe image, resulting in a reduction
of image FSV for the moist area. We address an approach to relate this parameter to the moisture content. 
We follow a three-step process: model development,
evaluation, and application. The developed model is rigorously
assessed and subsequently applied to estimate real leakage. 
The model development is performed according to the steps outlined in Fig.\ref{main_flowchart}. 
We place soil bags, each with a specified
moisture content, above the pipe. Fig. \ref{one_pipe2} provides a visual
representation of the pipe with an applied moist soil bag. The backscattered data of the proposed structure is captured on a specified number of sampling points upon step-frequency continuous wave (SFCW) modulation over a scan line perpendicularly positioned concerning the pipeline.  The raw data undergoes zero timing followed by clutter reduction and the pre-processed data is exploited to 
generate the images corresponding to various SMs, from which the FSV is calculated. 

We incorporate multiple frequency bands for each SM to enhance data diversity, contributing to diverse leakage dimensions, given the normalized-to-wavelength
parameter in the imaging algorithm. The acquired data is splitted up into two sets for model estimation (approximately
$70\%$ of the total data) and model evaluation (around $30\%$ of the total data), with various combinations considered. A
statistical fitting function is established for FSV regarding SM, and the probability distribution of fitting parameters is derived. Subsequently, the constructed model estimates leakage rates in
a practical scenario. In a laboratory-scale model mimicking SDI
system, a PVC pipe buried in the soil is sealed at both ends,
with water entry and exit facilitated through narrow hoses,
and a small hole is introduced on the pipe. By sequentially
capturing images of the leaking pipe and determining the FSV
at each moment, SM is measured using the developed model.
Differentiating SM over time allows for the determination of
the leakage rate. Fig.\ref{main_flowchart} outline the various steps of the  model parametrization and estimation.


\section{Image formation}\label{image formation}
As mentioned before, we need the moist region image to get the SM using the proposed approach.  Here,  we employ an inverse scattering and a synthetic focusing method to create the subsurface images and compare their performance. The former is based on a linearized scattering operator subject to the Born approximation. This results in a matrix equation that is solved by exploiting a regularization scheme due to the ill-posedness of the operator. While the Born approximation criteria are not met for the current application (Born Approximation Algorithm(BAA)  is applicable for the weak and small scatterers), it has been shown that it provides satisfactory results of  shape and size  even for the strong scatterers.     In this research, we explore the Back Projection Algorithm (BPA) as a synthetic focusing algorithm. BPA is simpler and more robust to measurement errors than BAA and it is more flexible specially for scenarios without explicit information about scatterers.  Before utilizing the image formation algorithm, it is necessary to eliminate the reflective effects from the ground, connectors, and couplings between antennas recognized as clutter. 
A clutter reduction technique is employed based on  SVD of B-scan matrix for the current problem.    
In this well-known method, the B-scan matrix is first decomposed using SVD, and then by eliminating  one or two eigenvectors corresponding to the largest eigenvalues, the clutter impact is removed. In the following we discuss the general aspects of two afformentioned algorithms for image formation.


\subsection{Born Approximation Algorithm}

The Born Approximation Algorithm(BAA) represents a widely recognized qualitative microwave imaging method detailed here. In accordance with Maxwell's equations, the scattered field expressed in terms of the equivalent contrast source is described by

\begin{equation}\label{BA}
    E_{s}(\Bar{r}_s,\omega) = -j\frac{\omega^2\mu_0\varepsilon_0}{4\pi} \int_S (\varepsilon(\Bar{r})-1) E_{i}(\Bar{r},\omega) H^{(2)}_0({k_s|\Bar{r}-\Bar{r_s}|)} d\Bar{r}
\end{equation}
under the zero-order Born approximation.
Here, $\varepsilon(\Bar{r})$ represents the target permittivity, and $H_0^{(2)}(.)$ denotes the zero-order Hankel function of the second kind.  Also, $\Bar{r}$ and $\Bar{r}_s$  stand for the position vector of scatterer and $s^{th}$ sampling point on the scan line, respectively.  
Considering this formulation as the data equation for an inverse scattering problem, the left-hand side of the equation is formulated based on the SFCW data acquired by a VNA at various antenna positions along the scan line. We aim to determine the unknown object function ($\tau=\varepsilon(\Bar{r})-1$), where the non-zero contrast function relies on the presence of a scatterer.
To discretize the equation, we mesh the reconstruction domain, approximating the integration over each mesh with a closed-form expression provided by \cite{Richmond}. Consequently, the matrix representation of \eqref{BA} is given by
\begin{equation}\label{BMF}
   \left[ E_{s}\right ]_{N\times 1} \approx \left [\Gamma \right ]_{N\times M}  \left [\tau \right ]_{M\times 1},
\end{equation}
where $N$ and $M$ denote the number of acquired data ($N=N_fN_s$, $N_f$ and $N_s$ respectively denote the number of frequencies and the number of sampling points on the scan line. ) and the number of meshes in the reconstruction domain, respectively.
The matrix entery at $n^{th}$ row and $m^{th}$ column, $\Gamma_{nm}$, is given by

\begin{equation}\label{Gamm}
\Gamma_{nm} = -aj\pi\frac{k_s}{2}  E_{i}(r_s^{(m)},\omega)J_1(k_s a) H_0^{(2)}(k_s r_s^{(m)})
\end{equation}
 where $k_s$ is the wave number in the soil for each frequency and $r_s^{(m)}$ is the distance of $m^{th}$ mesh to the measurement point.  Here, $a$ stands for the radius of an equivalent circle having the same area as the meshes.  
Here, $J_1(.)$ is the first-order Bessel function.
 $ E_{i}(r_s^{(m)},\omega)$ also specifies the incident wave approximated as a planar wave given by
\begin{equation}\label{E_inc}
     E_{i}(r_s^{(m)},\omega) = e^{-jk_s r_s^{(m)}}.
\end{equation}
The solution to \eqref{BMF} could be written as 
\begin{equation}\label{H_pinv}
    \left [\tau \right ]_{M\times 1}= \left [\Gamma \right ]^{-1}_{M\times N} \left[ E_{s}\right ]_{N\times 1}
\end{equation}
where $\left [\Gamma \right ]^{-1}_{M\times N}$  is the  pseudo-inverse of
$\left [\Gamma \right ]_{N\times M}$  constructed by the Truncated Singular Value Decomposition (TSVD) method given by 
\begin{equation}\label{H_pinv}
    \left [\Gamma \right ]^{-1}_{M\times N} =\left [ V\right ]_{M\times M}\left [ {\Sigma^{TSVD}}\right ]^{-1}_{M\times N}\left [U\right ]_{N\times N}^*.
\end{equation}
$\left [{\Sigma^{TSVD}}\right ]^{-1}_{M\times N}$ represents the inverse of the singular value matrix of $\left [ \Gamma \right ]$, where the Singular Values (SVs) less than a threshold are discarded before taking the matrix inverse.  Here, the columns of $\left [U\right ]_{N\times N}$ and $\left [U\right ]_{M\times M}$ depict the left and right singular vectors of $[\Gamma]$ respectively, and $[.]^*$ denotes the conjugate transpose of the matrix. We detail the threshold selection strategy for this algorithm in the simulation results section.
\subsection{Back Projection Algorithm  }

The next imaging approach employed in this research is BPA discussed here. BPA is based on refocusing of the scattered field achieved by compensating the phase shift of round-trip wave from the transmitting antenna to each pixel within the reconstruction domain and subsequently backscattering towards the receiving antenna. To implement BPA for an SFCW system, we need the scattered field for all scanning points and frequencies, to give the retrieved image by
\begin{equation}\label{back_pro_equ} \scalebox{0.85}{$
S(\Bar{r}_i) = \sum_{f = 1}^{N_f}\sum_{s = 1}^{N_s}(E_s(\Bar{r}_s),\omega_f) e^{2j(k_s^{(f)} |(\Bar{r}_i-\Bar{r}_s)|)}
$}.
\end{equation}
In this equation, $E_s(\Bar{r}_s,\omega_f)$ represents the measured scattered field, and $e^{2j(k_s^{(f)} |(\Bar{r}_i-\Bar{r}_s)|)}$ serves as phased compensating term according to signal path length through the soil. Here, $\Bar{r}_i$ and $\Bar{r}_s$  stand for the position vector of $i^{th}$ mesh and $s^{th}$ sampling point on the scan line, respectively and $k_s^{(f)}=\omega_f \sqrt{\varepsilon(\Bar{r}_i)\varepsilon_0\mu_0)}$ denotes the soil wavenumber for $f^{th}$ frequency, $\omega_f$ .
The expression given by \ref{back_pro_equ} ensures constructive summation of back-projected signals at locations of scatterers and destructive interference in areas devoid of scatterers within the reconstructed image. Here, we ignore the signal path through air since the antenna box is located at the closest vicinity of the soil surface.

\section{Scenarios \& results}\label{Senarios}
In this section, we explore the proposed approach outlined in Fig.\ref{main_flowchart} to develop and evaluate the model for the local SM estimation in an SDI system. 
We proceed to specify the SM for a laboratory-realized  SDI scenario through the obtained model. We also investigate the impact of plant root and soil pebbles by  considering the corresponding scenarios. 
\subsection{Laboratory Setup}
\begin{figure*}[htb]
	\centering
	\subfloat[]{\label{}\includegraphics[width=0.6\textwidth]{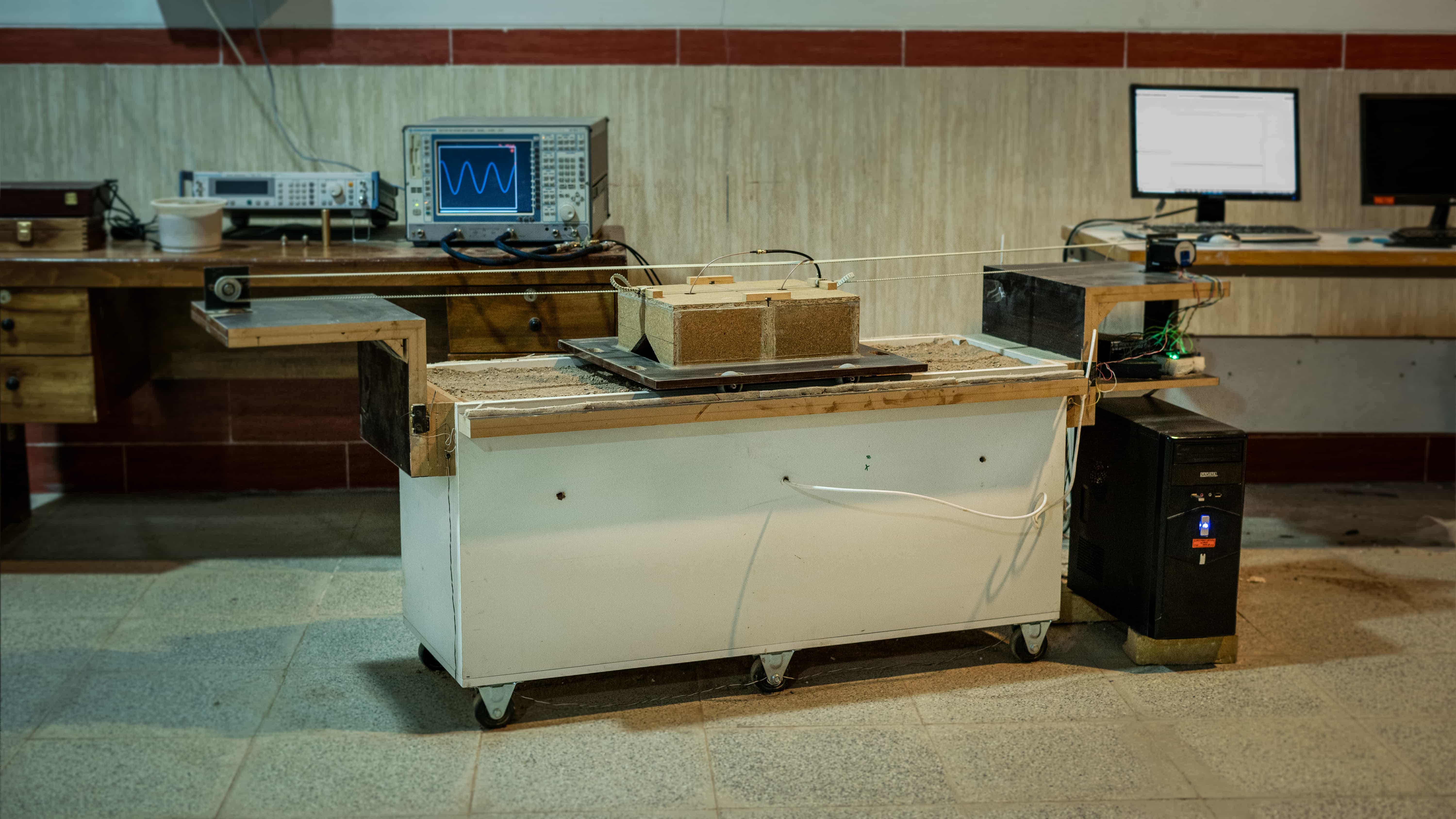}}
	\hfill
	\subfloat[]{\label{}\includegraphics[width=0.35\textwidth]{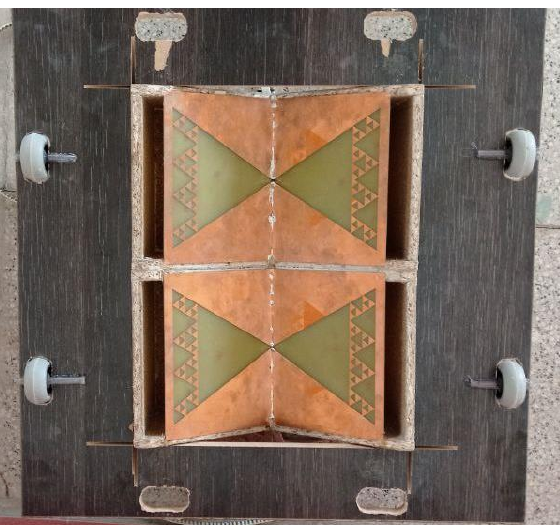}}
	\caption{(a) An overview of laboratory  setup including the solid test bed, VNA, antenna moving fixtures and processing tool. (b) TX and RX   bow-tie fractal antennas employed  for a wideband data acquisition}
	\label{scanning setup}
\end{figure*}

\begin{figure}[htb]
	\centering
	\includegraphics[scale=0.24, trim=0.5cm 0 0 2.5cm, clip]{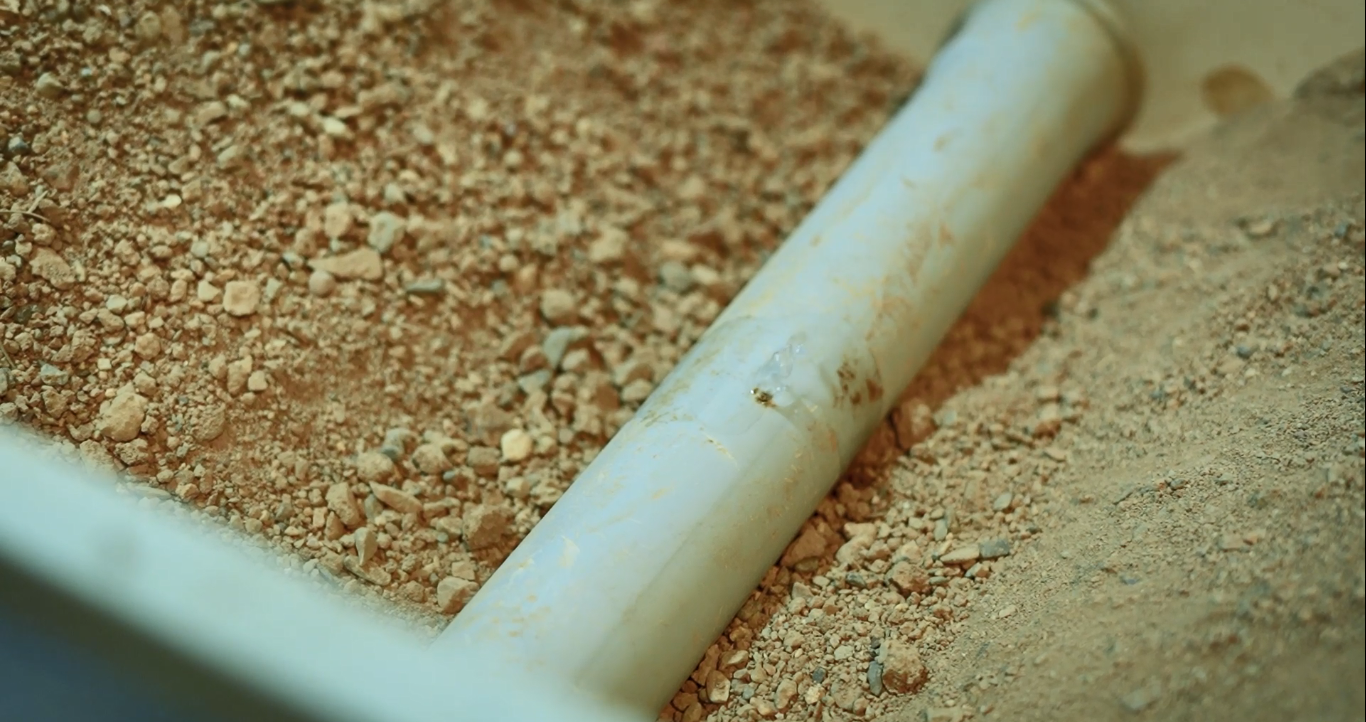}
	\caption{\small Leaking PVC pipe simulating an SDI structure.}
	\label{real_leak2}
\end{figure}

To assess the proposed method using practical data, we established a laboratory setup as depicted in Fig \ref{scanning setup}. 
The setup consists of a wooden enclosure measuring $\SI{120}{cm}$ in length, $\SI{40}{cm}$ in width, and $\SI{60}{cm}$ in height, filled with soft soil to accommodate a buried PVC pipe. This PVC pipe, resembling SDI pipes, is sealed at both ends using PVC caps and connected to thin plastic hoses for water inlet from a graduated valve-fitted bucket towards outlet.
Additionally, to simulate a real leakage scenario, a small hole was created on the PVC pipe facing upwards. Fig. \ref{real_leak2} illustrates structure for the drip simulation.

The data acquisition process involves utilizing a Vector Network Analyzer (VNA) operating at frequencies up to \SI{8}{GHz}, connected to a computer. Two bent  bow-tie fractal structure embedded into an enclosure serve  as transmitting and receiving antennas  and a \SI{21}{dB}-gain Low Noise Amplifier(LNA) is placed before the transmitting antenna to enhance the transmitter power. The values for the testbed and hardware parameters are given in Tab.\ref{tab:mytable}. 
\begin{table}[htbp]
	\centering
	\caption{The list of setup hardware and data acquisition parameters}
	\label{tab:mytable}
	\begin{tabular}{cccc}
		\toprule
		Parameters & Symbol & Value & Unit \\
		\midrule
		Transmitter power&  $P_t$ &  $15$ & $dBm$ \\
		 Length of scan line &  $L$ &  $1.2$ &  $m$ \\
		 Frequency band &  $f$ &  $1.2-3.775$ &  $GHz$ \\
		 Frequency step &  $\Delta f$ &  $25$ &  $MHz$ \\
		 No. data acquisition points &  $N_s$ &  $45$ &   \\
		 Tx \& Rx antenna Gain &  $G_t \& G_r$ &  $7$ &  $dB$ \\
		 LNA gain      &  $G_{LNA}$ &  $21$ &  $dB$ \\
		 Scan duration &  $T$     &  $14$ &  $min$ \\
          Pipe diameter & $D$    &   $4.5$ & $cm$ \\
          Pipe depth    & $d$ &$12$   &$cm$\\
		\bottomrule
	\end{tabular}
\end{table}
Automation of scanning is achieved by installing wheels beneath the antennas and creating a closed path using a timing belt connected to the antenna enclosure.  Movement of the antennas enclosure is facilitated by a stepper motor controlled by an Arduino, with MATLAB serving as the processing tool(Fig.\ref{scanning setup}).

\begin{figure}[htb]
	\centering	
	\input{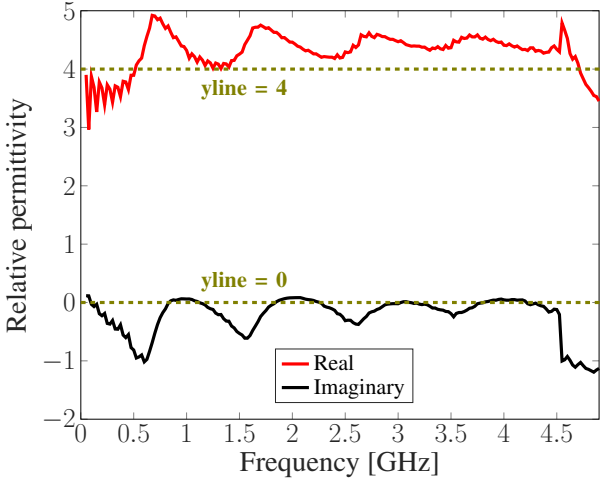}
	\caption{\small The real and imaginary parts of the measured soil relative permittivity.}
	\label{soil_eps}
\end{figure}

To generate a subsurface image, measurement of soil permittivity is essential to determine the wavenumber within the soil. Our investigation involves utilizing a transmission line method, wherein a coaxial structure filled with a soil sample is employed. $S_{21}(\omega)$ is measured for this structure and correlated with soil permittivity through a closed-form expression given by \cite{coax_model}.
Fig.\ref{soil_eps} shows the estimated permittivity of the applied soil for the frequencies spanning from $0$ to \SI{5}{GHz} covering the data acquisition band.
As shown, the real and imaginary parts of the permittivity are approximated to $4$ and $0$, respectively throughout the band.

\subsection{Model development and evaluation results}
\begin{figure}[htb]
\centering
\input{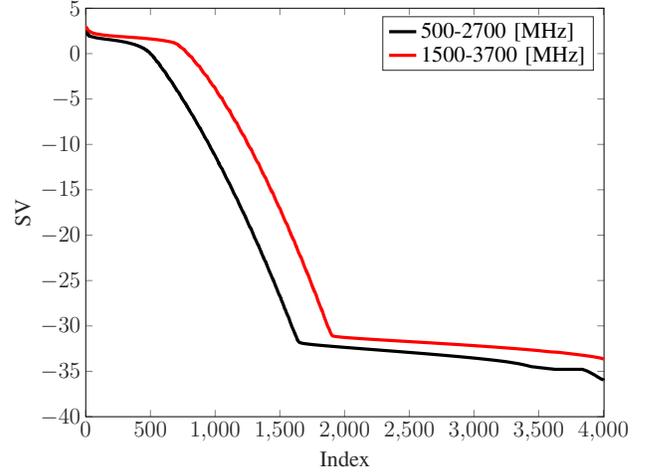}
\caption{\small Singular value pattern of $[\Gamma]$ for two frequency bands.}
\label{SV}
\end{figure}

 System setup parameters for the data capturing are given in Tab. \ref{tab:mytable}.  We incorporate moist soil bags of eight different  SM percentages from  $12.5\%$ to $100\%$ with the step of $12.5\%$. 
Fig. \ref{SV} depicts the singular values pattern of $[\Gamma]$ given by \eqref{Gamm} for two frequency bands. The figure clearly demonstrates that the higher frequencies offer richer information since they provide more SVs for the signal subspace of $[\Gamma]$. The figure is also  exploited for threshold specification of TSVD applied for  BAA realization as  mentioned previously.  In this context, the index at which a sharp drop in the SV curve occurs is selected as the threshold value distinguishing the signal and noise Subspace of $[\Gamma]$ above which the SVs are truncated.
We also emphasize that the clutter removal is accomplished via the SVD technique with elimination of one singular value.

\begin{figure*}[htb]
	\centering	
	\input{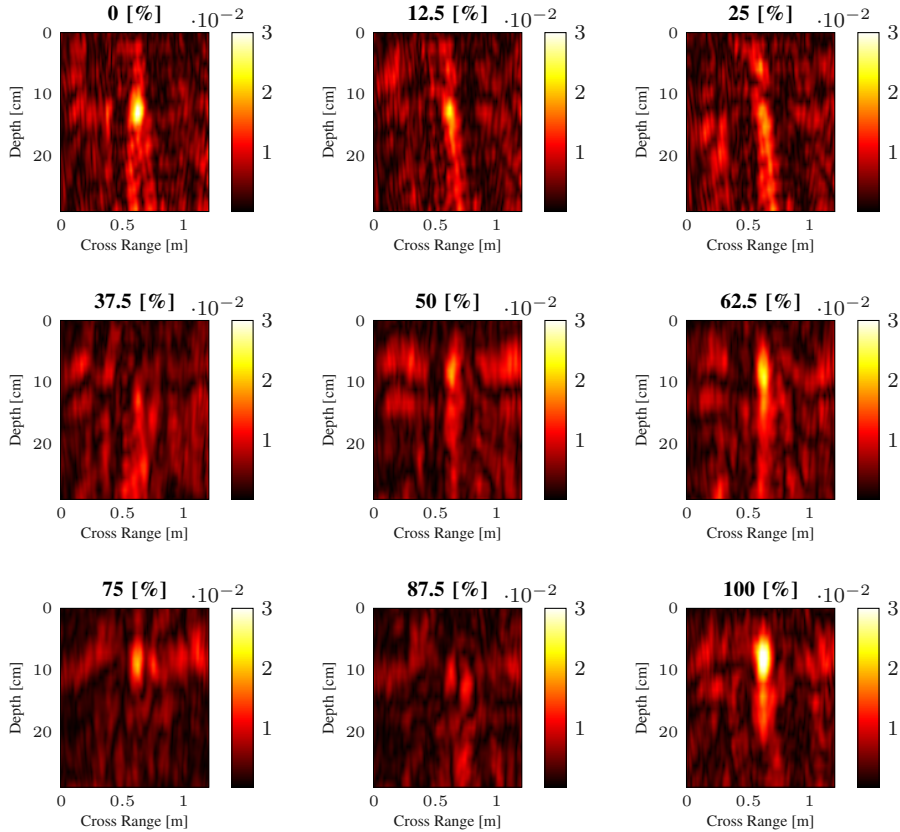}
	\caption{\small The retrieved image of a single pipe in the presence of leakage model bags at various moisture percentages using the BAA method.}
	\label{SM_real_leakـ_Born}
\end{figure*}

\begin{figure}[htb]
	\centering	
	\input{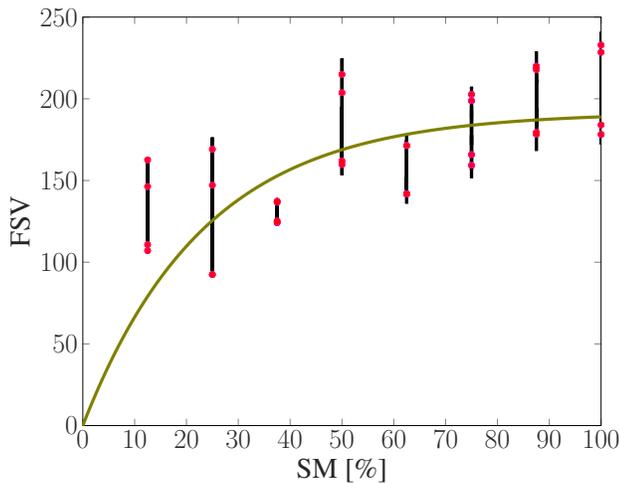}
	\caption{\small Measured and exponentially fitted FSV  for $16$ frequency bands across $8$ SMs by the proposed approach outlined in \ref{fig:flowchart}.}
	\label{curve_estimat}
\end{figure}

Figs.\ref{SM_real_leakـ_Born} and \ref{SM_real_leak_Back}  provide  BAA and BPA images corresponding to these SM percentages within the frequency range of \SI{1.3}{GHz}-\SI{3.5}{GHz} after zero timing and clutter removal.
The figure encompasses the buried pipe in dry soil, enabling precise determination of its location and dimensions. This confirms the efficacy 
\begin{figure*}[htb]
	\centering
	
	\input{wet_packet.tex}
	\caption{\small The retrieved image of a single pipe in the presence of leakage model bags at various moisture percentages using the BPA) method.}
	\label{SM_real_leak_Back}
\end{figure*}
It's noticeable that as the SM increases, the visibility of the pipe diminishes gradually while the signal return from the moist area intensifies. Additionally, Fig.\ref{curve_estimat} exhibits the FSV of the reconstructed images within the moist area across $8$ SM percentages and $16$ overlapped frequency bands spanning \SI{1.3}{GHz}-\SI{3.5}{GHz}  with a bandwidth of \SI{2.2}{GHz}.  The starting frequencies of each band are spaced by \SI{25}{MHz}. The figure illustrates a consistent upward trend in FSV for both BPA and BAA images.

To provide data diversity encompassing various electrical length of moisture extend,  we utilize a combination of $12$ frequency bands, representing approximately $70\%$ of the total $16$ frequency bands, resulting in 1820 data points for model development.
Each set of $12$ frequency bands yields an FSV result, which is then fitted with an exponential function given by
\begin{equation}
    FSV=a\left(1-e^{b\times SM}\right)
\end{equation}
and the histograms of estimated parameters $a,b$ are given by Fig. \ref{born_hist} for BAA and Fig. \ref{back_pro_hist} for BPA.  
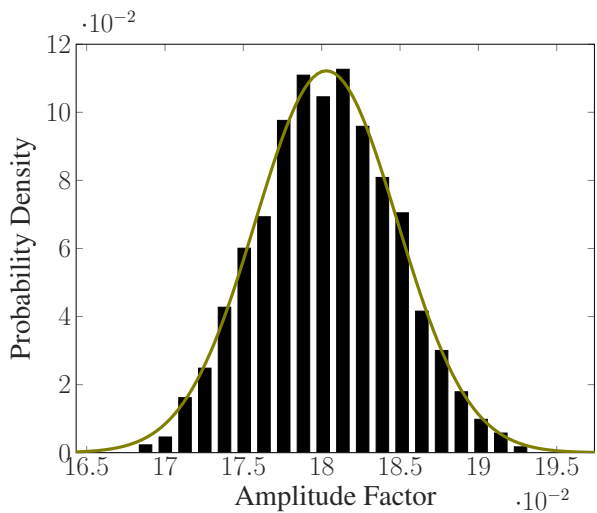
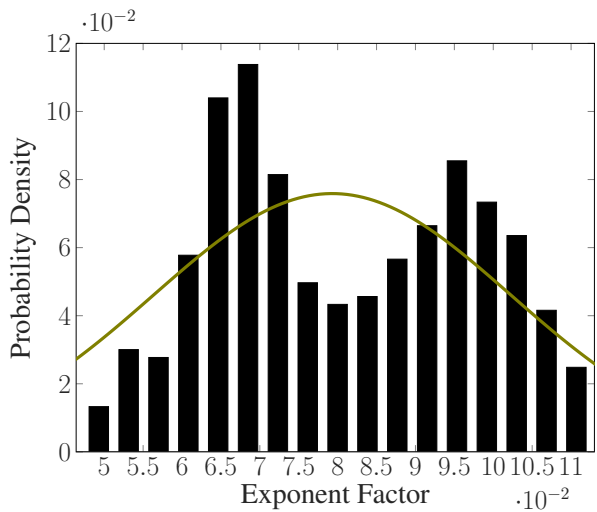
\begin{figure}[tb]
	
	\subfloat[]{
%
%
\definecolor{mycolor1}{rgb}{0.00000,0.44700,0.74100}%
\definecolor{mycolor2}{rgb}{1.00000,0.00000,1.00000}%
\definecolor{mycolor3}{rgb}{0.50196, 0.50196, 0.00000}
\begin{tikzpicture}[scale=0.6]

\begin{axis}[%
 scale only axis,
width=4.521in,
height=3.566in,
at={(0.758in,0.481in)},
scale only axis,
bar shift auto,
xmin=0.16433,
xmax=0.197405,
xticklabel style={font=\bfseries\fontsize{16}{16}\selectfont, color=white!15!black},
xlabel style={font=\fontsize{18}{18}\selectfont, color=white!15!black},
xlabel={Amplitude Factor},
ymin=0,
ymax=0.12,
yticklabel style={font=\bfseries\fontsize{16}{16}\selectfont, color=white!15!black},
ylabel style={font=\fontsize{18}{18}\selectfont, color=white!15!black},
ylabel={Probability Density},
scaled x ticks=base 10:2,
scaled y ticks=base 10:2,
axis background/.style={fill=white}
]
\addplot[ybar, bar width=0.0008, fill=black, draw=black, area legend] table[row sep=crcr] {%
0.168759841152439	0.00231080300404391\\
0.17001932780609	0.00462160600808781\\
0.17127881445974	0.0161756210283073\\
0.172538301113391	0.024841132293472\\
0.173797787767042	0.0427498555748122\\
0.175057274420692	0.0600808781051415\\
0.176316761074343	0.0693240901213172\\
0.177576247727993	0.097631426920855\\
0.178835734381644	0.110918544194107\\
0.180095221035294	0.104563835932987\\
0.181354707688945	0.11265164644714\\
0.182614194342596	0.0958983246678221\\
0.183873680996246	0.0808781051415367\\
0.185133167649897	0.0704794916233391\\
0.186392654303547	0.0415944540727903\\
0.187652140957198	0.0300404390525708\\
0.188911627610849	0.0179087232813403\\
0.190171114264499	0.0098209127671866\\
0.19143060091815	0.00577700751010976\\
0.1926900875718	0.00173310225303293\\
};
\addplot[forget plot, color=white!15!black] table[row sep=crcr] {%
0.16433	0\\
0.197405	0\\
};
\addplot [color=mycolor3, line width=2.0pt, forget plot]
  table[row sep=crcr]{%
0.163759841152439	0.000141316574018785\\
0.16404496927361	0.000177544809025291\\
0.164330097394781	0.000222177158336055\\
0.164615225515953	0.000276928329690739\\
0.164900353637124	0.000343804783064862\\
0.165185481758295	0.00042514100244486\\
0.165470609879466	0.000523637354210713\\
0.165755738000637	0.000642398912625471\\
0.166040866121808	0.00078497445542435\\
0.166325994242979	0.00095539464114266\\
0.16661112236415	0.00115820817894663\\
0.166896250485321	0.00139851459707504\\
0.167181378606492	0.00168199201476463\\
0.167466506727663	0.00201491813336651\\
0.167751634848835	0.00240418249525456\\
0.168036762970006	0.00285728792522707\\
0.168321891091177	0.00338233898042218\\
0.168607019212348	0.00398801520378288\\
0.168892147333519	0.00468352701527442\\
0.16917727545469	0.00547855219619582\\
0.169462403575861	0.00638315113555906\\
0.169747531697032	0.00740765932211361\\
0.170032659818203	0.00856255598682763\\
0.170317787939375	0.0098583083306009\\
0.170602916060546	0.0113051914084762\\
0.170888044181717	0.012913084477552\\
0.171173172302888	0.0146912454387279\\
0.171458300424059	0.0166480658942753\\
0.17174342854523	0.0187908102802881\\
0.172028556666401	0.021125343486272\\
0.172313684787572	0.0236558523095616\\
0.172598812908743	0.026384566972132\\
0.172883941029914	0.029311489711218\\
0.173169069151086	0.0324341381013129\\
0.173454197272257	0.0357473112325592\\
0.173739325393428	0.03924288712076\\
0.174024453514599	0.0429096597233653\\
0.17430958163577	0.0467332236566419\\
0.174594709756941	0.0506959141332918\\
0.174879837878112	0.0547768087588848\\
0.175164965999283	0.0589517966433283\\
0.175450094120454	0.063193718816589\\
0.175735222241625	0.0674725822156438\\
0.176020350362797	0.0717558475747451\\
0.176305478483968	0.0760087894582286\\
0.176590606605139	0.0801949244896363\\
0.17687573472631	0.0842765016266179\\
0.177160862847481	0.0882150461874751\\
0.177445990968652	0.0919719473344114\\
0.177731119089823	0.095509076941712\\
0.178016247210994	0.0987894263007284\\
0.178301375332165	0.101777746006\\
0.178586503453336	0.10444117368468\\
0.178871631574508	0.106749834016528\\
0.179156759695679	0.108677395768853\\
0.17944188781685	0.110201571345702\\
0.179727015938021	0.111304545609495\\
0.180012144059192	0.111973322442577\\
0.180297272180363	0.112199979623553\\
0.180582400301534	0.111981825029759\\
0.180867528422705	0.111321449859781\\
0.181152656543876	0.110226677404881\\
0.181437784665047	0.1087104087854\\
0.181722912786219	0.106790369905083\\
0.18200804090739	0.104488766562687\\
0.182293169028561	0.101831857102212\\
0.182578297149732	0.0988494540975205\\
0.182863425270903	0.0955743682845213\\
0.183148553392074	0.0920418092219873\\
0.183433681513245	0.0882887579464773\\
0.183718809634416	0.0843533271734446\\
0.184003937755587	0.0802741243909787\\
0.184289065876759	0.0760896325190498\\
0.18457419399793	0.07183762170723\\
0.184859322119101	0.0675546043739755\\
0.185144450240272	0.0632753438187506\\
0.185429578361443	0.0590324247410265\\
0.185714706482614	0.0548558918585256\\
0.185999834603785	0.0507729606131954\\
0.186284962724956	0.0468078017668459\\
0.186570090846127	0.0429813995931278\\
0.186855218967298	0.0393114814340499\\
0.18714034708847	0.035812514662501\\
0.187425475209641	0.0324957656200213\\
0.187710603330812	0.0293694139112787\\
0.187995731451983	0.0264387145499762\\
0.188280859573154	0.0237061998691668\\
0.188565987694325	0.0211719128242846\\
0.188851115815496	0.0188336633112196\\
0.189136243936667	0.0166872993674855\\
0.189421372057838	0.0147269855881556\\
0.189706500179009	0.0129454817313686\\
0.189991628300181	0.0113344152698028\\
0.190276756421352	0.00988454252294981\\
0.190561884542523	0.00858599393974927\\
0.190847012663694	0.00742850005433763\\
0.191132140784865	0.00640159557533347\\
0.191417268906036	0.00549479996203928\\
0.191702397027207	0.0046977736653814\\
0.191987525148378	0.00400044994914014\\
0.192272653269549	0.00339314284550316\\
0.19255778139072	0.00286663233100957\\
0.192842909511892	0.00241222823226172\\
0.193128037633063	0.00202181468738102\\
0.193413165754234	0.00168787720467792\\
0.193698293875405	0.00140351448282051\\
0.193983421996576	0.00116243719741327\\
0.194268550117747	0.000958955929140638\\
0.194553678238918	0.000787960320933927\\
0.194838806360089	0.000644891418471944\\
0.19512393448126	0.000525708981746078\\
0.195409062602431	0.000426855366579405\\
0.195694190723603	0.00034521737389991\\
0.195979318844774	0.000278087259904787\\
0.196264446965945	0.000223123899246648\\
0.196549575087116	0.000178314901792154\\
0.196834703208287	0.000141940305699574\\
0.197119831329458	0.000112538308567324\\
0.197404959450629	8.8873356068376e-05\\
0.1976900875718	6.99067846300428e-05\\
};
\end{axis}

\begin{axis}[%
width=5.833in,
height=4.375in,
at={(0in,0in)},
scale only axis,
xmin=0,
xmax=1,
ymin=0,
ymax=1,
axis line style={draw=none},
ticks=none,
axis x line*=bottom,
axis y line*=left
]
\end{axis}
\end{tikzpicture}%
}
	
	\subfloat[]{
%
%
\definecolor{mycolor1}{rgb}{0.00000,0.44700,0.74100}%
\definecolor{mycolor2}{rgb}{1.00000,0.00000,1.00000}%
\definecolor{mycolor3}{rgb}{0.50196, 0.50196, 0.00000}
\begin{tikzpicture}[scale = 0.6]

\begin{axis}[%
 scale only axis,
width=4.521in,
height=3.566in,
at={(0.758in,0.481in)},
scale only axis,
bar shift auto,
xmin=0.0463918,
xmax=0.113,
xticklabel style={font=\bfseries\fontsize{16}{16}\selectfont, color=white!15!black},
xlabel style={font=\fontsize{18}{18}\selectfont, color=white!15!black},
xlabel={Exponent Factor},
ymin=0,
ymax=0.12,
yticklabel style={font=\bfseries\fontsize{16}{16}\selectfont, color=white!15!black},
ylabel style={font=\fontsize{18}{18}\selectfont, color=white!15!black},
ylabel={Probability Density},
scaled x ticks=base 10:2,
scaled y ticks=base 10:2,
axis background/.style={fill=white}
]
\addplot[ybar, bar width=0.0025, fill=black, draw=black, area legend] table[row sep=crcr] {%
0.0493021475906232	0.0132871172732525\\
0.0531384038896487	0.0300404390525708\\
0.0569746601886742	0.0277296360485269\\
0.0608109164876997	0.0577700751010976\\
0.0646471727867252	0.103986135181976\\
0.0684834290857507	0.113807047949162\\
0.0723196853847762	0.0814558058925477\\
0.0761559416838018	0.049682264586944\\
0.0799921979828273	0.0433275563258232\\
0.0838284542818528	0.0456383593298671\\
0.0876647105808783	0.0566146735990757\\
0.0915009668799038	0.0664355863662623\\
0.0953372231789293	0.0854997111496245\\
0.0991734794779548	0.073367995378394\\
0.10300973577698	0.0635470826112074\\
0.106845992076006	0.0415944540727903\\
0.110682248375031	0.024841132293472\\
0.114518504674057	0.0132871172732525\\
0.118354760973082	0.00577700751010976\\
0.122191017272108	0.00231080300404391\\
};
\addplot[forget plot, color=white!15!black] table[row sep=crcr] {%
0.0463918	0\\
0.125101	0\\
};
\addplot [color=mycolor3, line width=2.0pt, forget plot]
  table[row sep=crcr]{%
0.0443021475906232	0.0238133282948325\\
0.044998692713997	0.0249261318998373\\
0.0456952378373708	0.0260669937847663\\
0.0463917829607446	0.0272350563388346\\
0.0470883280841185	0.028429346557173\\
0.0477848732074923	0.0296487745218708\\
0.0484814183308661	0.0308921323637482\\
0.0491779634542399	0.0321580937317658\\
0.0498745085776137	0.0334452137948193\\
0.0505710537009876	0.0347519297982224\\
0.0512675988243614	0.0360765621944761\\
0.0519641439477352	0.0374173163649513\\
0.052660689071109	0.0387722849459031\\
0.0533572341944828	0.0401394507687989\\
0.0540537793178567	0.0415166904212985\\
0.0547503244412305	0.0429017784314044\\
0.0554468695646043	0.044292392073313\\
0.0561434146879781	0.0456861167893932\\
0.056839959811352	0.0470804522185021\\
0.0575365049347258	0.0484728188165751\\
0.0582330500580996	0.0498605650511134\\
0.0589295951814734	0.0512409751468819\\
0.0596261403048472	0.0526112773558601\\
0.0603226854282211	0.0539686527202871\\
0.0610192305515949	0.0553102442935587\\
0.0617157756749687	0.056633166779794\\
0.0624123207983425	0.0579345165491397\\
0.0631088659217164	0.0592113819823497\\
0.0638054110450902	0.0604608540949065\\
0.064501956168464	0.0616800373879685\\
0.0651985012918378	0.062866060870772\\
0.0658950464152116	0.0640160891968056\\
0.0665915915385855	0.0651273338541531\\
0.0672881366619593	0.0661970643488711\\
0.0679846817853331	0.067222619319171\\
0.0686812269087069	0.0682014175175103\\
0.0693777720320807	0.0691309685974947\\
0.0700743171554546	0.0700088836427477\\
0.0707708622788284	0.0708328853756281\\
0.0714674074022022	0.0716008179848731\\
0.072163952525576	0.072310656512904\\
0.0728604976489498	0.0729605157456488\\
0.0735570427723237	0.073548658550307\\
0.0742535878956975	0.0740735036094715\\
0.0749501330190713	0.0745336325034352\\
0.0756466781424451	0.0749277960962956\\
0.076343223265819	0.0752549201856242\\
0.0770397683891928	0.07551411037994\\
0.0777363135125666	0.0757046561729927\\
0.0784328586359404	0.0758260341888806\\
0.0791294037593142	0.0758779105772591\\
0.0798259488826881	0.0758601425432988\\
0.0805224940060619	0.0757727790025794\\
0.0812190391294357	0.0756160603567168\\
0.0819155842528095	0.0753904173911627\\
0.0826121293761833	0.0750964693022504\\
0.0833086744995572	0.0747350208661324\\
0.084005219622931	0.0743070587677266\\
0.0847017647463048	0.0738137471131064\\
0.0853983098696786	0.0732564221539022\\
0.0860948549930525	0.0726365862571713\\
0.0867914001164263	0.0719559011588198\\
0.0874879452398001	0.0712161805429702\\
0.0881844903631739	0.0704193819936388\\
0.0888810354865477	0.0695675983686876\\
0.0895775806099216	0.0686630486492119\\
0.0902741257332954	0.0677080683203066\\
0.0909706708566692	0.0667050993414922\\
0.091667215980043	0.0656566797669683\\
0.0923637611034168	0.0645654330772872\\
0.0930603062267907	0.0634340572849936\\
0.0937568513501645	0.0622653138772637\\
0.0944533964735383	0.0610620166585941\\
0.0951499415969121	0.0598270205561551\\
0.095846486720286	0.0585632104495292\\
0.0965430318436598	0.0572734900852448\\
0.0972395769670336	0.0559607711347733\\
0.0979361220904074	0.0546279624525408\\
0.0986326672137812	0.0532779595880129\\
0.0993292123371551	0.051913634603085\\
0.100025757460529	0.0505378262428764\\
0.100722302583903	0.0491533305046219\\
0.101418847707277	0.0477628916457017\\
0.10211539283065	0.0463691936680065\\
0.102811937954024	0.0449748523118106\\
0.103508483077398	0.043582407588181\\
0.104205028200772	0.042194316874708\\
0.104901573324146	0.0408129485950478\\
0.105598118447519	0.0394405764984518\\
0.106294663570893	0.038079374551158\\
0.106991208694267	0.0367314124472739\\
0.107687753817641	0.0353986517426165\\
0.108384298941015	0.0340829426109251\\
0.109080844064389	0.0327860212179611\\
0.109777389187762	0.0315095077052696\\
0.110473934311136	0.0302549047718447\\
0.11117047943451	0.0290235968386102\\
0.111867024557884	0.0278168497775449\\
0.112563569681258	0.0266358111844382\\
0.113260114804631	0.0254815111716943\\
0.113956659928005	0.0243548636552973\\
0.114653205051379	0.0232566681080375\\
0.115349750174753	0.0221876117493618\\
0.116046295298127	0.0211482721407708\\
0.116742840421501	0.0201391201545234\\
0.117439385544874	0.0191605232825401\\
0.118135930668248	0.0182127492517938\\
0.118832475791622	0.0172959699121541\\
0.119529020914996	0.0164102653625781\\
0.12022556603837	0.0155556282817244\\
0.120922111161744	0.0147319684294753\\
0.121618656285117	0.0139391172864881\\
0.122315201408491	0.013176832799726\\
0.123011746531865	0.0124448042029435\\
0.123708291655239	0.0117426568822852\\
0.124404836778613	0.0110699572584942\\
0.125101381901986	0.0104262176586941\\
0.12579792702536	0.00981090115228467\\
0.126494472148734	0.00922342632716351\\
0.127191017272108	0.00866317198422841\\
};
\end{axis}

\begin{axis}[%
width=5.833in,
height=4.375in,
at={(0in,0in)},
scale only axis,
xmin=0,
xmax=1,
xticklabel style={font=\bfseries\fontsize{14}{14}\selectfont, color=white!15!black},
xlabel style={font=\fontsize{14}{14}\selectfont, color=white!15!black},
ymin=0,
ymax=1,
yticklabel style={font=\bfseries\fontsize{14}{14}\selectfont, color=white!15!black},
ylabel style={font=\fontsize{14}{14}\selectfont, color=white!15!black},
axis line style={draw=none},
ticks=none,
axis x line*=bottom,
axis y line*=left
]
\end{axis}
\end{tikzpicture}%
}
	
	\caption{Histogram and fitted Gaussian PDF for the (a) amplitude and (b) exponent factors of exponentially fitting  of a set of $1820$ data points of measured FSV considering BAA.}
	\label{born_hist}
\end{figure}
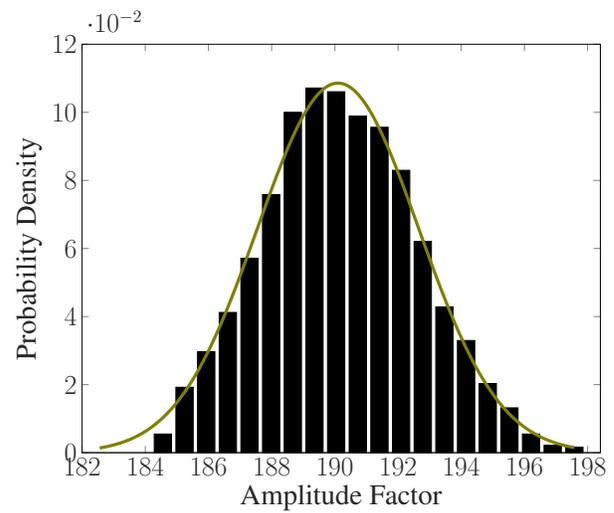
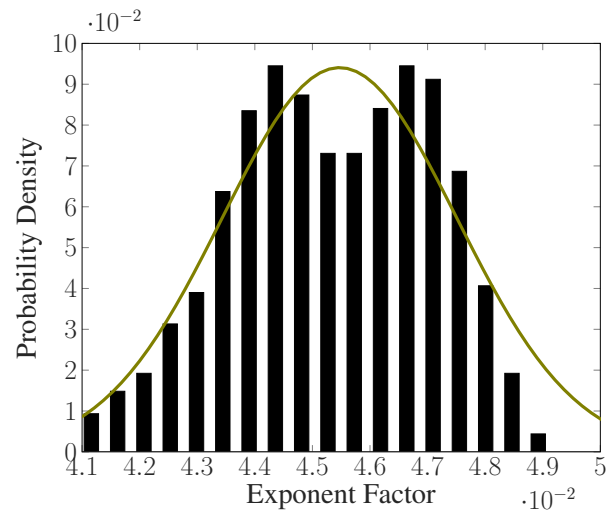
\begin{figure}[tb]
	
	\subfloat[]{
%
%
\definecolor{mycolor1}{rgb}{0.00000,0.44700,0.74100}%
\definecolor{mycolor2}{rgb}{1.00000,0.00000,1.00000}%
\definecolor{mycolor3}{rgb}{0.50196, 0.50196, 0.00000}
\begin{tikzpicture}[scale = 0.6]

\begin{axis}[%
 scale only axis,
width=4.521in,
height=3.566in,
at={(0.758in,0.481in)},
scale only axis,
bar shift auto,
xmin=182,
xmax=198.403054828713,
xticklabel style={font=\bfseries\fontsize{16}{16}\selectfont, color=white!15!black},
xlabel style={font=\fontsize{18}{18}\selectfont, color=white!15!black},
xlabel={Amplitude Factor},
ymin=0,
ymax=0.12,
yticklabel style={font=\bfseries\fontsize{16}{16}\selectfont, color=white!15!black},
ylabel style={font=\fontsize{18}{18}\selectfont, color=white!15!black},
ylabel={Probability Density},
scaled y ticks=base 10:2,
axis background/.style={fill=white}
]
\addplot[ybar, bar width=0.548, fill=black, draw=black, area legend] table[row sep=crcr] {%
184.557316135145	0.00549450549450549\\
185.242748743737	0.0192307692307692\\
185.92818135233	0.0296703296703297\\
186.613613960922	0.0412087912087912\\
187.299046569515	0.0571428571428571\\
187.984479178107	0.0758241758241758\\
188.6699117867	0.1\\
189.355344395292	0.107142857142857\\
190.040777003885	0.106043956043956\\
190.726209612477	0.0989010989010989\\
191.41164222107	0.0956043956043956\\
192.097074829662	0.082967032967033\\
192.782507438255	0.0620879120879121\\
193.467940046847	0.0428571428571429\\
194.15337265544	0.032967032967033\\
194.838805264032	0.0203296703296703\\
195.524237872625	0.0131868131868132\\
196.209670481217	0.00549450549450549\\
196.89510308981	0.0021978021978022\\
197.580535698402	0.00164835164835165\\
};
\addplot[forget plot, color=white!15!black] table[row sep=crcr] {%
182	0\\
198.403054828713	0\\
};
\addplot [color=mycolor3, line width=2.0pt, forget plot]
  table[row sep=crcr]{%
182.557316135145	0.00138007394609659\\
182.683561677693	0.00159528151510039\\
182.809807220241	0.00183954343147456\\
182.93605276279	0.00211602354629511\\
183.062298305338	0.00242811182094227\\
183.188543847886	0.00277942274766139\\
183.314789390435	0.00317379059636298\\
183.441034932983	0.00361526109475944\\
183.567280475532	0.00410807915684855\\
183.69352601808	0.00465667229184398\\
183.819771560628	0.0052656293528336\\
183.946017103177	0.00593967432250143\\
184.072262645725	0.00668363488277075\\
184.198508188273	0.00750240557661424\\
184.324753730822	0.00840090544365047\\
184.45099927337	0.00938403009635873\\
184.577244815919	0.0104565983002858\\
184.703490358467	0.0116232932286664\\
184.829735901015	0.0128885986782022\\
184.955981443564	0.0142567306567437\\
185.082226986112	0.0157315648833158\\
185.208472528661	0.0173165608739449\\
185.334718071209	0.019014683420394\\
185.460963613757	0.0208283224001581\\
185.587209156306	0.0227592119816452\\
185.713454698854	0.0248083504049082\\
185.839700241402	0.0269759216219874\\
185.965945783951	0.0292612201682949\\
186.092191326499	0.0316625807039631\\
186.218436869048	0.0341773137083207\\
186.344682411596	0.0368016488286275\\
186.470927954144	0.0395306873731498\\
186.597173496693	0.0423583653965153\\
186.723419039241	0.0452774287504211\\
186.84966458179	0.0482794213644204\\
186.975910124338	0.0513546878795362\\
187.102155666886	0.0544923915826975\\
187.228401209435	0.057680548384098\\
187.354646751983	0.060906077345088\\
187.480892294531	0.0641548680046436\\
187.60713783708	0.0674118644720226\\
187.733383379628	0.0706611659570909\\
187.859628922177	0.0738861431037178\\
187.985874464725	0.0770695691819372\\
188.112120007273	0.0801937648880919\\
188.238365549822	0.0832407552058715\\
188.36461109237	0.086192436502238\\
188.490856634919	0.0890307517776725\\
188.617102177467	0.0917378717666831\\
188.743347720015	0.0942963793983634\\
188.869593262564	0.0966894549834097\\
188.995838805112	0.0989010593981466\\
189.12208434766	0.100916112491352\\
189.248329890209	0.102720663948511\\
189.374575432757	0.104302053911815\\
189.500820975306	0.105649060772615\\
189.627066517854	0.106752033724767\\
189.753312060402	0.107603007889475\\
189.879557602951	0.108195800090954\\
190.005803145499	0.10852608367196\\
190.132048688047	0.10859144108292\\
190.258294230596	0.108391393350493\\
190.384539773144	0.107927405923083\\
190.510785315693	0.107202870793514\\
190.637030858241	0.106223065203949\\
190.763276400789	0.104995087636296\\
190.889521943338	0.103527772174173\\
191.015767485886	0.101831582681618\\
191.142013028435	0.0999184885715047\\
191.268258570983	0.097801824226299\\
191.394504113531	0.095496134379419\\
191.52074965608	0.0930170079624854\\
191.646995198628	0.0903809030686303\\
191.773240741176	0.0876049657727741\\
191.899486283725	0.0847068455855252\\
192.025731826273	0.0817045102987598\\
192.151977368822	0.0786160629098534\\
192.27822291137	0.0754595631910103\\
192.404468453918	0.0722528563044246\\
192.530713996467	0.0690134106581386\\
192.656959539015	0.065758166957391\\
192.783205081564	0.062503400138467\\
192.909450624112	0.0592645955834079\\
193.03569616666	0.0560563407115984\\
193.161941709209	0.0528922327352521\\
193.288187251757	0.0497848030571294\\
193.414432794305	0.0467454584870408\\
193.540678336854	0.0437844391648299\\
193.666923879402	0.0409107928071249\\
193.793169421951	0.0381323646478106\\
193.919414964499	0.0354558022218299\\
194.045660507047	0.0328865739515838\\
194.171906049596	0.0304290003369164\\
194.298151592144	0.0280862964247477\\
194.424397134693	0.0258606241431076\\
194.550642677241	0.0237531530262399\\
194.676888219789	0.021764127831307\\
194.803133762338	0.0198929415512308\\
194.929379304886	0.0181382123598627\\
195.055624847434	0.0164978630821262\\
195.181870389983	0.0149692018598032\\
195.308115932531	0.0135490027797271\\
195.43436147508	0.0122335853417813\\
195.560607017628	0.0110188917656075\\
195.686852560176	0.00990056126381114\\
195.813098102725	0.00887400054231229\\
195.939343645273	0.00793444992217093\\
196.065589187822	0.0070770446088728\\
196.19183473037	0.00629687076213074\\
196.318080272918	0.00558901613958547\\
196.444325815467	0.00494861519958152\\
196.570571358015	0.00437088865003102\\
196.696816900563	0.00385117752125663\\
196.823062443112	0.00338497191994096\\
196.94930798566	0.00296793468860356\\
197.075553528209	0.00259592025035211\\
197.201799070757	0.00226498896229612\\
197.328044613305	0.00197141733345403\\
197.454290155854	0.00171170448493758\\
197.580535698402	0.00148257524250682\\
};
\end{axis}

\begin{axis}[%
width=5.833in,
height=4.375in,
at={(0in,0in)},
scale only axis,
xmin=0,
xmax=1,
ymin=0,
ymax=1,
axis line style={draw=none},
ticks=none,
axis x line*=bottom,
axis y line*=left
]
\end{axis}
\end{tikzpicture}%
}
	
	\subfloat[]{
%
%
\definecolor{mycolor1}{rgb}{0.00000,0.44700,0.74100}%
\definecolor{mycolor2}{rgb}{1.00000,0.00000,1.00000}%
\definecolor{mycolor3}{rgb}{0.50196, 0.50196, 0.00000}
\begin{tikzpicture}[scale=0.6]

\begin{axis}[%
 scale only axis,
width=4.521in,
height=3.566in,
at={(0.758in,0.481in)},
scale only axis,
bar shift auto,
xmin=4.1e-2,
xmax=0.05,
xticklabel style={font=\bfseries\fontsize{16}{16}\selectfont, color=white!15!black},
xlabel={Exponent Factor},
xlabel style={font=\fontsize{18}{18}\selectfont, color=white!15!black},
ymin=0,
ymax=0.1,
yticklabel style={font=\bfseries\fontsize{16}{16}\selectfont, color=white!15!black},
ylabel={Probability Density},
ylabel style={font=\fontsize{18}{18}\selectfont, color=white!15!black},
scaled x ticks=base 10:2,
scaled y ticks=base 10:2,
axis background/.style={fill=white}
]
\addplot[ybar, bar width=0.00025, fill=black, draw=black, area legend] table[row sep=crcr] {%
0.0402481833063737	0.00274725274725275\\
0.0407046934004058	0.00549450549450549\\
0.0411612034944379	0.00934065934065934\\
0.0416177135884701	0.0148351648351648\\
0.0420742236825022	0.0192307692307692\\
0.0425307337765344	0.0313186813186813\\
0.0429872438705665	0.039010989010989\\
0.0434437539645987	0.0637362637362637\\
0.0439002640586308	0.0835164835164835\\
0.044356774152663	0.0945054945054945\\
0.0448132842466951	0.0873626373626374\\
0.0452697943407273	0.0730769230769231\\
0.0457263044347594	0.0730769230769231\\
0.0461828145287916	0.0840659340659341\\
0.0466393246228237	0.0945054945054945\\
0.0470958347168559	0.0912087912087912\\
0.047552344810888	0.0686813186813187\\
0.0480088549049202	0.0406593406593407\\
0.0484653649989523	0.0192307692307692\\
0.0489218750929844	0.0043956043956044\\
};
\addplot[forget plot, color=white!15!black] table[row sep=crcr] {%
0.0393379	0\\
0.05	0\\
};
\addplot [color=mycolor3, line width=2.0pt, forget plot]
  table[row sep=crcr]{%
0.0352481833063737	3.50062286339308e-07\\
0.0354051050860931	5.12346079591355e-07\\
0.0355620268658125	7.45456705500365e-07\\
0.0357189486455319	1.07825715756724e-06\\
0.0358758704252513	1.5504693648408e-06\\
0.0360327922049707	2.21638368837914e-06\\
0.0361897139846902	3.14968858712799e-06\\
0.0363466357644096	4.44970483988963e-06\\
0.036503557544129	6.2493623584767e-06\\
0.0366604793238484	8.72531522896433e-06\\
0.0368174011035678	1.2110650216543e-05\\
0.0369743228832873	1.67107026496276e-05\\
0.0371312446630067	2.29225473130414e-05\\
0.0372881664427261	3.12587753977216e-05\\
0.0374450882224455	4.23761949574287e-05\\
0.0376020100021649	5.71100935750855e-05\\
0.0377589317818844	7.65146685626534e-05\\
0.0379158535616038	0.00010191015139852\\
0.0380727753413232	0.000134937017889317\\
0.0382296971210426	0.000177617472212308\\
0.038386618900762	0.000232424110677053\\
0.0385435406804814	0.000302355300542036\\
0.0387004624602009	0.000391016344249703\\
0.0388573842399203	0.000502704938050044\\
0.0390143060196397	0.000642498780069934\\
0.0391712277993591	0.00081634244773885\\
0.0393281495790785	0.00103112986817958\\
0.039485071358798	0.0012947778777131\\
0.0396419931385174	0.00161628554859364\\
0.0397989149182368	0.00200577320366719\\
0.0399558366979562	0.00247449440380039\\
0.0401127584776756	0.00303481374759213\\
0.040269680257395	0.00370014314205257\\
0.0404266020371145	0.00448482936162074\\
0.0405835238168339	0.00540398628198508\\
0.0407404455965533	0.00647326621542374\\
0.0408973673762727	0.00770856632975389\\
0.0410542891559921	0.00912566822381181\\
0.0412112109357116	0.010739811348887\\
0.041368132715431	0.0125652040622322\\
0.0415250544951504	0.0146144795905669\\
0.0416819762748698	0.0168981079419574\\
0.0418388980545892	0.0194237786670023\\
0.0419958198343086	0.0221957731324873\\
0.0421527416140281	0.0252143484023546\\
0.0423096633937475	0.0284751576751331\\
0.0424665851734669	0.0319687342549687\\
0.0426235069531863	0.035680067000794\\
0.0427804287329057	0.0395882949036227\\
0.0429373505126252	0.0436665467352971\\
0.0430942722923446	0.0478819485113413\\
0.043251194072064	0.0521958168163258\\
0.0434081158517834	0.0565640499448604\\
0.0435650376315028	0.0609377215037193\\
0.0437219594112222	0.06526387288298\\
0.0438788811909417	0.0694864922026668\\
0.0440358029706611	0.0735476584096313\\
0.0441927247503805	0.0773888206146752\\
0.0443496465300999	0.0809521750153466\\
0.0445065683098193	0.0841820953226573\\
0.0446634900895388	0.0870265679288544\\
0.0448204118692582	0.0894385804671814\\
0.0449773336489776	0.091377412164911\\
0.045134255428697	0.0928097765911749\\
0.0452911772084164	0.0937107720233851\\
0.0454480989881359	0.0940646015279192\\
0.0456050207678553	0.0938650336620512\\
0.0457619425475747	0.093115585023658\\
0.0459188643272941	0.0918294171764417\\
0.0460757861070135	0.0900289521721012\\
0.0462327078867329	0.0877452223627292\\
0.0463896296664524	0.0850169808465882\\
0.0465465514461718	0.0818896081707434\\
0.0467034732258912	0.0784138583640536\\
0.0468603950056106	0.0746444926477\\
0.04701731678533	0.0706388520557664\\
0.0471742385650495	0.0664554206270733\\
0.0473311603447689	0.0621524288763412\\
0.0474880821244883	0.0577865431254467\\
0.0476450039042077	0.0534116802954766\\
0.0478019256839271	0.0490779803369075\\
0.0479588474636465	0.0448309600754929\\
0.048115769243366	0.0407108633672868\\
0.0482726910230854	0.036752213572482\\
0.0484296128028048	0.0329835659226624\\
0.0485865345825242	0.0294274497566114\\
0.0487434563622436	0.0261004841430257\\
0.0489003781419631	0.0230136453097445\\
0.0490572999216825	0.0201726606776677\\
0.0492142217014019	0.017578502178784\\
0.0493711434811213	0.0152279508614899\\
0.0495280652608407	0.01311420541992\\
0.0496849870405601	0.0112275090384581\\
0.0498419088202796	0.00955577159099332\\
0.049998830599999	0.00808516753021033\\
0.0501557523797184	0.00680069349628689\\
0.0503126741594378	0.0056866735306462\\
0.0504695959391572	0.00472720358777914\\
0.0506265177188767	0.00390653061943141\\
0.0507834394985961	0.00320936472275866\\
0.0509403612783155	0.00262112560081179\\
0.0510972830580349	0.00212812682370252\\
0.0512542048377543	0.00171770308276825\\
0.0514111266174737	0.00137828681055908\\
0.0515680483971932	0.00109944123456148\\
0.0517249701769126	0.000871857199001918\\
0.051881891956632	0.000687320995543152\\
0.0520388137363514	0.000538660064611285\\
0.0521957355160708	0.000419672839207962\\
0.0523526572957903	0.000325048273057092\\
0.0525095790755097	0.000250279788228677\\
0.0526665008552291	0.000191577548010139\\
0.0528234226349485	0.000145782152331862\\
0.0529803444146679	0.000110282098408675\\
0.0531372661943873	8.29366711826711e-05\\
0.0532941879741068	6.20053402527478e-05\\
0.0534511097538262	4.60842480635089e-05\\
0.0536080315335456	3.40499777274634e-05\\
0.053764953313265	2.50104827464998e-05\\
0.0539218750929844	1.82628365630637e-05\\
};
\end{axis}

\begin{axis}[%
width=5.833in,
height=4.375in,
at={(0in,0in)},
scale only axis,
xmin=0,
xmax=1,
ymin=0,
ymax=1,
axis line style={draw=none},
ticks=none,
axis x line*=bottom,
axis y line*=left
]
\end{axis}
\end{tikzpicture}%
}
	
	\caption{Histogram and fitted Gaussian PDF for the (a) amplitude and (b) exponent factors of exponentially fitting  of a set of $1820$ data points of measured FSV considering BPA.}
	\label{back_pro_hist}
\end{figure}
It's evident that both the amplitude factor, $a$, and the exponent factor, $b$, conform well to Gaussian Probability Distribution Functions (PDF),whose mean values and variances specified in Tab. \ref{ab charactiriz}.
\begin{table}[htbp]
	\centering
	\caption{Statistical parameters of Gaussian PDF fitted to the amplitude and exponent factors}
	\label{ab charactiriz}
	\begin{tabular}{ccc}
		\toprule
		& BPA & BAA  \\
		\midrule
		$\sigma^2_a$& 0.0015 &  $2.0486 \times 10^{-5}$ \\
		$\sigma^2_b$ &  $4.1789 \times 10^{-6}$ &  $5.2845 \times 10^{-4}$  \\
		$\mu_a$&  0.05 &  0.1803  \\
		$\mu_b$ &  0.04547&  0.0793 \\
		\bottomrule
	\end{tabular}
\end{table}  
To assess the performance of the developed model, we utilize the remaining $4$ bands (approximately $30\%$ of the total frequency bands) for each set of $12$ bands. The SM Estimation Error (SMEE) is then computed as 
\begin{equation}
    SMEE = \overline{SM^{EST}-SM^{EXT}}
\end{equation}
Here, $SM^{EST}$ and $SM^{EXT}$ represent the estimated and exact values respectively, and $\overline{(.)}$ stands for averaging over  $4$ bands. The results are illustrated in Fig. \ref{CR_2k} for each SM.
For BPA, the SMEE typically ranges around $10\%$, with lower errors observed for lower moisture scenarios, and potentially increasing up to $30\%$ for higher moisture scenarios. Notably, BPA demonstrates lower errors, particularly for higher moisture content, compared to BAA, where errors can reach up to $70\%$. We've also generated SMEE results  when the clutter reduction is performed by discarding two SVs of B-sacn matrix rather than one SV, depicted in Fig. \ref{error_CR_3k}. The figure highlights a notable increase in SMEE, particularly for BAA. This underscores the importance of the second SV of B-scan matrix, which contains significant information about the SM content.

\begin{figure}[tb]
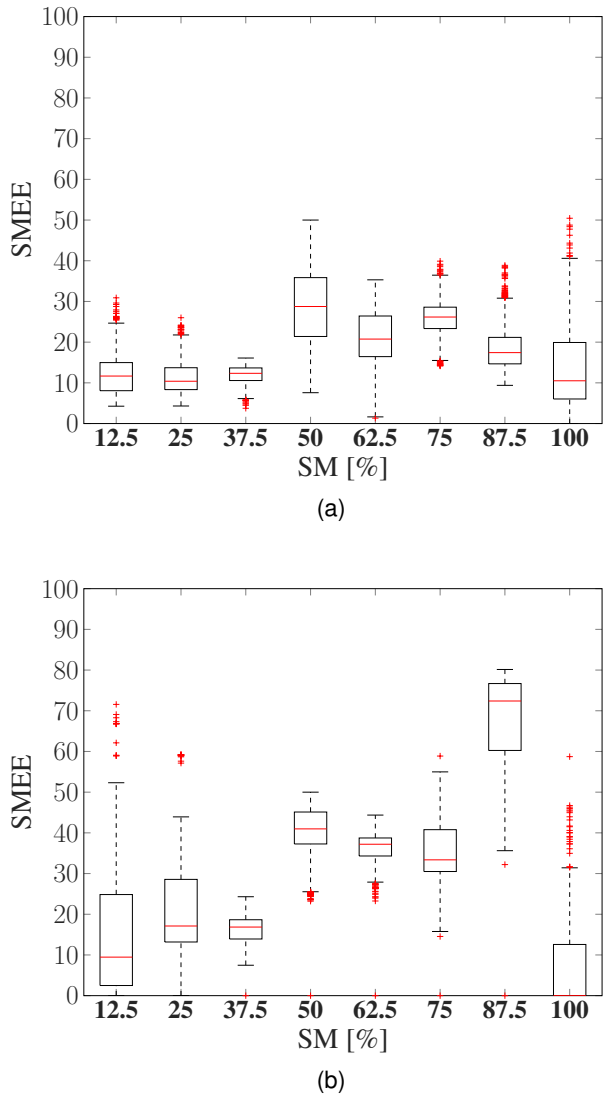

	
	\subfloat[]{\input{back_pro_error.tikz}}
	
	\subfloat[]{\input{Born_error.tikz}}
	
	\caption{Model evaluation by SMEE illustrated for (a) BPA and (b) BAA displaying BPA outperforming specially for higher SM values. }
	\label{CR_2k}
\end{figure}

\begin{figure}[tb]
	
	\subfloat[]{
%
%
\begin{tikzpicture}[scale=0.6]

\begin{axis}[%
width=4.521in,
height=3.553in,
at={(0.758in,0.494in)}, 
scale only axis,
xmin=0.5,
xmax=8.5,
xtick={1,2,3,4,5,6,7,8},
xticklabels={12.5, 25, 37.5, 50, 62.5, 75, 87.5, 100},
xticklabel style={font=\bfseries\fontsize{16}{16}\selectfont, color=white!15!black},
xlabel={SM [\%]},
xlabel style={font=\fontsize{18}{18}\selectfont, color=white!15!black},
ymin=0,
ymax=100,
yticklabel style={font=\bfseries\fontsize{16}{16}\selectfont, color=white!15!black},
ylabel={SMEE},
ylabel style={font=\fontsize{18}{18}\selectfont, color=white!15!black},
axis background/.style={fill=white}
]
\addplot [color=black, dashed, forget plot]
  table[row sep=crcr]{%
1	26.6688542170149\\
1	58.7073861600149\\
};
\addplot [color=black, dashed, forget plot]
  table[row sep=crcr]{%
2	28.7947538628212\\
2	44.0498595180617\\
};
\addplot [color=black, dashed, forget plot]
  table[row sep=crcr]{%
3	24.6793389953886\\
3	26.8008256623779\\
};
\addplot [color=black, dashed, forget plot]
  table[row sep=crcr]{%
4	40.209892364277\\
4	50.0958287391355\\
};
\addplot [color=black, dashed, forget plot]
  table[row sep=crcr]{%
5	43.5323723656395\\
5	50.1582632402486\\
};
\addplot [color=black, dashed, forget plot]
  table[row sep=crcr]{%
6	43.3782463971868\\
6	59.3637133958718\\
};
\addplot [color=black, dashed, forget plot]
  table[row sep=crcr]{%
7	39.1363936573832\\
7	62.5166052130948\\
};
\addplot [color=black, dashed, forget plot]
  table[row sep=crcr]{%
8	34.1448652855365\\
8	72.4796790938828\\
};
\addplot [color=black, dashed, forget plot]
  table[row sep=crcr]{%
1	0\\
1	4.18195845102647\\
};
\addplot [color=black, dashed, forget plot]
  table[row sep=crcr]{%
2	0\\
2	15.6731470597865\\
};
\addplot [color=black, dashed, forget plot]
  table[row sep=crcr]{%
3	4.8664374206215\\
3	16.7213642148419\\
};
\addplot [color=black, dashed, forget plot]
  table[row sep=crcr]{%
4	21.8061984910087\\
4	32.5725468896148\\
};
\addplot [color=black, dashed, forget plot]
  table[row sep=crcr]{%
5	29.8560925297233\\
5	38.0571914273237\\
};
\addplot [color=black, dashed, forget plot]
  table[row sep=crcr]{%
6	19.0314600154853\\
6	31.6904571596538\\
};
\addplot [color=black, dashed, forget plot]
  table[row sep=crcr]{%
7	0\\
7	23.4164853401444\\
};
\addplot [color=black, dashed, forget plot]
  table[row sep=crcr]{%
8	0\\
8	8.02887399057973\\
};
\addplot [color=black, forget plot]
  table[row sep=crcr]{%
0.875	58.7073861600149\\
1.125	58.7073861600149\\
};
\addplot [color=black, forget plot]
  table[row sep=crcr]{%
1.875	44.0498595180617\\
2.125	44.0498595180617\\
};
\addplot [color=black, forget plot]
  table[row sep=crcr]{%
2.875	26.8008256623779\\
3.125	26.8008256623779\\
};
\addplot [color=black, forget plot]
  table[row sep=crcr]{%
3.875	50.0958287391355\\
4.125	50.0958287391355\\
};
\addplot [color=black, forget plot]
  table[row sep=crcr]{%
4.875	50.1582632402486\\
5.125	50.1582632402486\\
};
\addplot [color=black, forget plot]
  table[row sep=crcr]{%
5.875	59.3637133958718\\
6.125	59.3637133958718\\
};
\addplot [color=black, forget plot]
  table[row sep=crcr]{%
6.875	62.5166052130948\\
7.125	62.5166052130948\\
};
\addplot [color=black, forget plot]
  table[row sep=crcr]{%
7.875	72.4796790938828\\
8.125	72.4796790938828\\
};
\addplot [color=black, forget plot]
  table[row sep=crcr]{%
0.875	0\\
1.125	0\\
};
\addplot [color=black, forget plot]
  table[row sep=crcr]{%
1.875	0\\
2.125	0\\
};
\addplot [color=black, forget plot]
  table[row sep=crcr]{%
2.875	4.8664374206215\\
3.125	4.8664374206215\\
};
\addplot [color=black, forget plot]
  table[row sep=crcr]{%
3.875	21.8061984910087\\
4.125	21.8061984910087\\
};
\addplot [color=black, forget plot]
  table[row sep=crcr]{%
4.875	29.8560925297233\\
5.125	29.8560925297233\\
};
\addplot [color=black, forget plot]
  table[row sep=crcr]{%
5.875	19.0314600154853\\
6.125	19.0314600154853\\
};
\addplot [color=black, forget plot]
  table[row sep=crcr]{%
6.875	0\\
7.125	0\\
};
\addplot [color=black, forget plot]
  table[row sep=crcr]{%
7.875	0\\
8.125	0\\
};
\addplot [color=black, forget plot]
  table[row sep=crcr]{%
0.75	4.18195845102647\\
0.75	26.6688542170149\\
1.25	26.6688542170149\\
1.25	4.18195845102647\\
0.75	4.18195845102647\\
};
\addplot [color=black, forget plot]
  table[row sep=crcr]{%
1.75	15.6731470597865\\
1.75	28.7947538628212\\
2.25	28.7947538628212\\
2.25	15.6731470597865\\
1.75	15.6731470597865\\
};
\addplot [color=black, forget plot]
  table[row sep=crcr]{%
2.75	16.7213642148419\\
2.75	24.6793389953886\\
3.25	24.6793389953886\\
3.25	16.7213642148419\\
2.75	16.7213642148419\\
};
\addplot [color=black, forget plot]
  table[row sep=crcr]{%
3.75	32.5725468896148\\
3.75	40.209892364277\\
4.25	40.209892364277\\
4.25	32.5725468896148\\
3.75	32.5725468896148\\
};
\addplot [color=black, forget plot]
  table[row sep=crcr]{%
4.75	38.0571914273237\\
4.75	43.5323723656395\\
5.25	43.5323723656395\\
5.25	38.0571914273237\\
4.75	38.0571914273237\\
};
\addplot [color=black, forget plot]
  table[row sep=crcr]{%
5.75	31.6904571596538\\
5.75	43.3782463971868\\
6.25	43.3782463971868\\
6.25	31.6904571596538\\
5.75	31.6904571596538\\
};
\addplot [color=black, forget plot]
  table[row sep=crcr]{%
6.75	23.4164853401444\\
6.75	39.1363936573832\\
7.25	39.1363936573832\\
7.25	23.4164853401444\\
6.75	23.4164853401444\\
};
\addplot [color=black, forget plot]
  table[row sep=crcr]{%
7.75	8.02887399057973\\
7.75	34.1448652855365\\
8.25	34.1448652855365\\
8.25	8.02887399057973\\
7.75	8.02887399057973\\
};
\addplot [color=red, forget plot]
  table[row sep=crcr]{%
0.75	22.7697707074145\\
1.25	22.7697707074145\\
};
\addplot [color=red, forget plot]
  table[row sep=crcr]{%
1.75	25.9012441885734\\
2.25	25.9012441885734\\
};
\addplot [color=red, forget plot]
  table[row sep=crcr]{%
2.75	19.8682690394365\\
3.25	19.8682690394365\\
};
\addplot [color=red, forget plot]
  table[row sep=crcr]{%
3.75	36.1472914966136\\
4.25	36.1472914966136\\
};
\addplot [color=red, forget plot]
  table[row sep=crcr]{%
4.75	40.4000752637235\\
5.25	40.4000752637235\\
};
\addplot [color=red, forget plot]
  table[row sep=crcr]{%
5.75	37.442715925267\\
6.25	37.442715925267\\
};
\addplot [color=red, forget plot]
  table[row sep=crcr]{%
6.75	29.2476252135963\\
7.25	29.2476252135963\\
};
\addplot [color=red, forget plot]
  table[row sep=crcr]{%
7.75	18.6860285832441\\
8.25	18.6860285832441\\
};
\addplot [color=black, only marks, mark=+, mark options={solid, draw=red}, forget plot]
  table[row sep=crcr]{%
1	66.9997048015651\\
1	67.0003354652683\\
1	67.0381264499098\\
1	67.1423832403612\\
1	67.1882360447573\\
1	67.2130432721993\\
1	67.5137928947728\\
1	67.615479014286\\
1	67.9128327289141\\
1	68.0259879701283\\
1	70.6424917540365\\
1	73.6535651835258\\
1	74.0594492891834\\
};
\addplot [color=black, only marks, mark=+, mark options={solid, draw=red}, forget plot]
  table[row sep=crcr]{%
2	56.7184631453205\\
2	57.7038467303373\\
2	57.7720472635677\\
2	57.7925476198761\\
2	57.8562686717513\\
2	58.0252994046864\\
2	58.0386329794345\\
2	58.193249413016\\
2	58.2724856920227\\
2	58.3405058112883\\
2	58.4380616790424\\
2	58.7447572367356\\
2	58.983322659876\\
};
\addplot [color=black, only marks, mark=+, mark options={solid, draw=red}, forget plot]
  table[row sep=crcr]{%
3	0\\
3	3.54952116668206\\
3	3.66091120659062\\
3	3.67969167879139\\
3	3.81116109757434\\
3	3.92013647802713\\
3	3.97612229246265\\
3	4.00839313479555\\
3	4.18508394915954\\
3	4.2808178692329\\
3	4.32796817712567\\
3	4.41212459230827\\
3	4.57717932110227\\
3	4.65001974738651\\
3	4.65386783283699\\
3	4.71489693902994\\
};
\addplot [color=black, only marks, mark=+, mark options={solid, draw=red}, forget plot]
  table[row sep=crcr]{%
4	0\\
4	53.9211381705772\\
4	55.9693510223818\\
};
\addplot [color=black, only marks, mark=+, mark options={solid, draw=red}, forget plot]
  table[row sep=crcr]{%
5	0\\
5	27.4194457916919\\
5	28.1280504214438\\
5	28.2382190495986\\
5	28.3281531326175\\
5	28.4339339315634\\
5	28.6229993515931\\
5	28.6623160089007\\
5	28.8392615723343\\
5	29.1274634697263\\
5	29.2132618967014\\
5	29.3555456995585\\
5	29.3946390077638\\
5	29.4548679381275\\
5	29.4622362095685\\
5	29.525932663716\\
5	53.2551249894836\\
5	53.5629057060608\\
};
\addplot [color=black, only marks, mark=+, mark options={solid, draw=red}, forget plot]
  table[row sep=crcr]{%
6	0\\
};
\addplot [color=black, only marks, mark=+, mark options={solid, draw=red}, forget plot]
  table[row sep=crcr]{%
7	63.116032770432\\
7	63.1552898250697\\
7	63.2211251179609\\
7	63.6491033272447\\
7	63.7154935585976\\
7	63.8554958795201\\
7	64.3081863063606\\
7	64.3359695163306\\
7	64.365586529721\\
7	64.3738752004322\\
7	64.8134812187523\\
7	64.9286368872021\\
7	65.3566208527193\\
7	65.3965076892037\\
7	65.5231832831738\\
7	65.9369362611129\\
7	65.969098349462\\
7	66.5056384576398\\
7	66.5101389037994\\
7	66.906233811984\\
7	66.9167227237637\\
7	66.9714832090106\\
7	67.3358083122938\\
7	67.3722772307027\\
7	67.4168780349017\\
7	67.8566901825421\\
7	69.2598841232899\\
};
\addplot [color=black, only marks, mark=+, mark options={solid, draw=red}, forget plot]
  table[row sep=crcr]{%
8	73.6686490197939\\
8	74.657346610834\\
8	75.4088649232275\\
8	75.7362445939494\\
8	76.8267298926094\\
8	78.4806938981415\\
};
\end{axis}

\begin{axis}[%
width=5.833in,
height=4.375in,
at={(0in,0in)},
scale only axis,
xmin=0,
xmax=1,
ymin=0,
ymax=1,
axis line style={draw=none},
ticks=none,
axis x line*=bottom,
axis y line*=left
]
\end{axis}
\end{tikzpicture}%
}
	
	\subfloat[]{\input{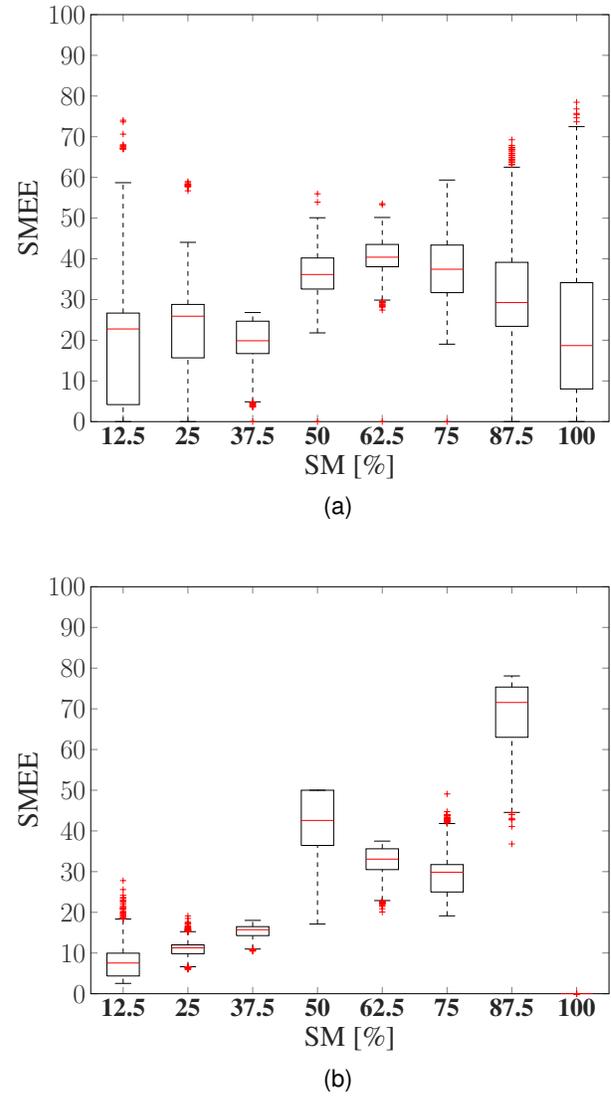}}
	
	\caption{Model evaluation by SMEE illustrated for (a) BPA and (b) BAA considering the SVD clutter reduction with 2 singular value  eliminations.}
	\label{error_CR_3k}
\end{figure}

\subsection{SM estimation in the presence of medium clutter}
\begin{figure}[tb]
	\centering
	\subfloat[]{s\includegraphics[width=0.22\textwidth]{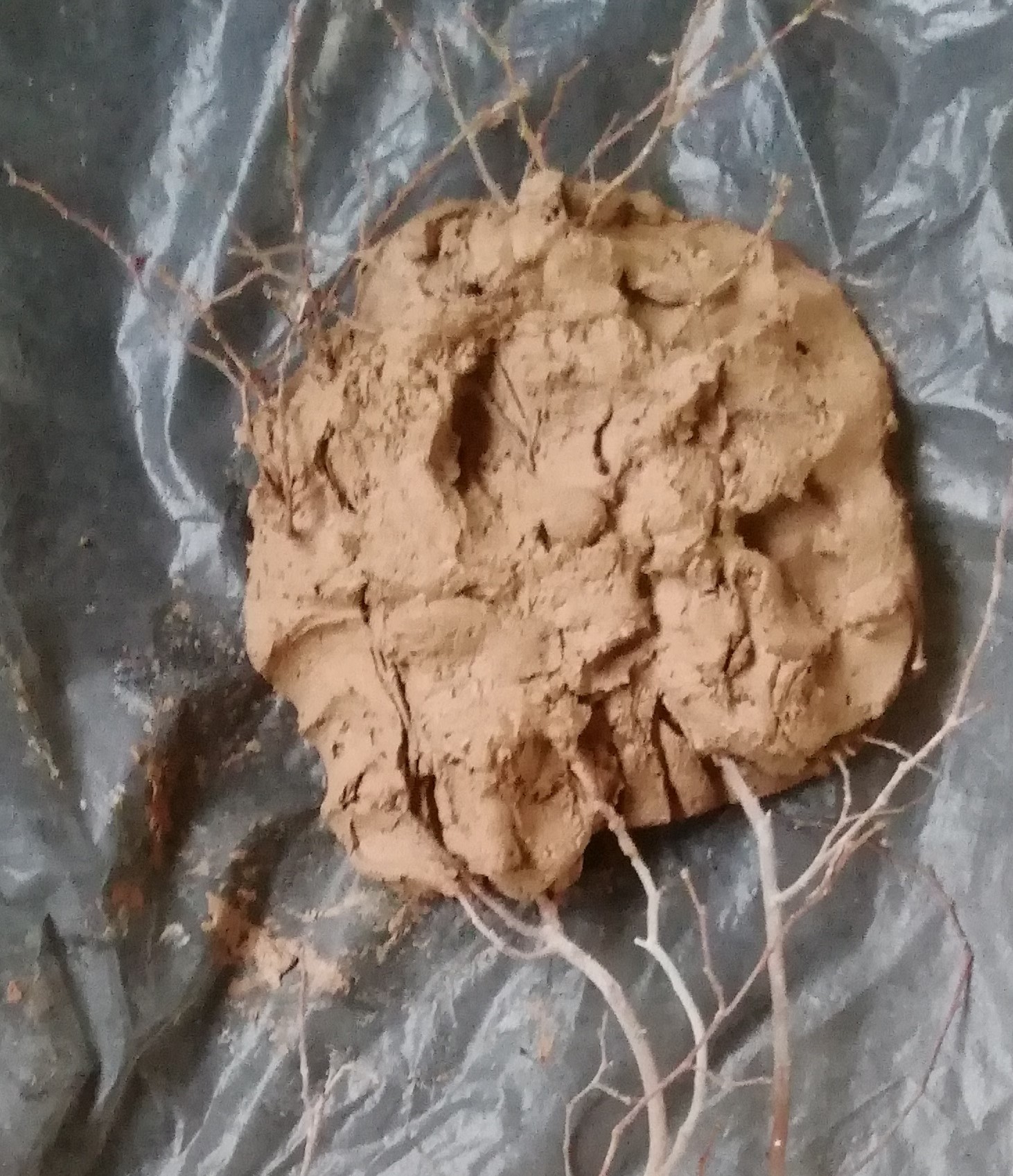}}
	\hfill
	\subfloat[]{\includegraphics[width=0.22\textwidth]{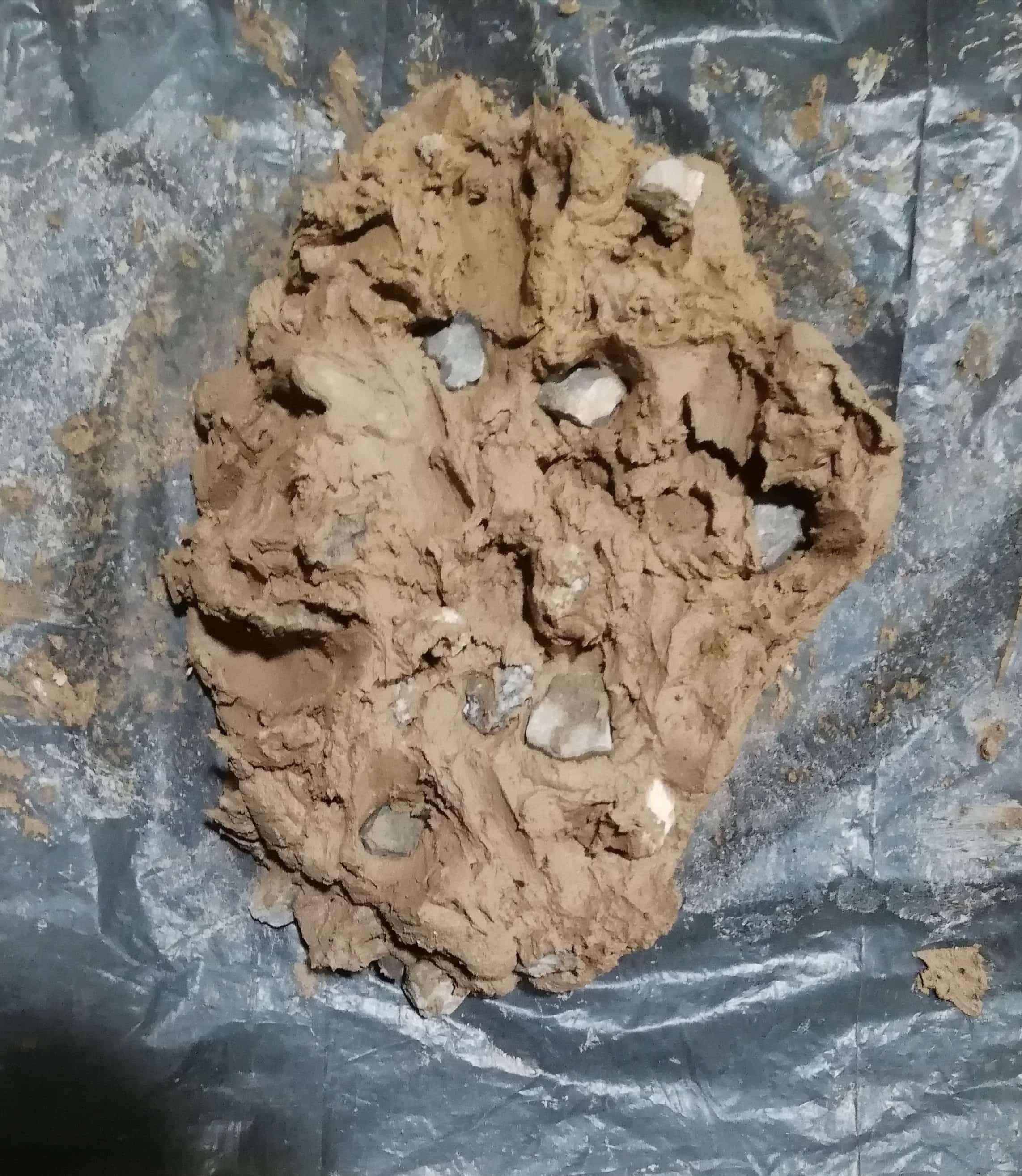}}
	\caption{Soil bag of $60\%$ moisture with added (a) roots and (b) pebbles.}
	\label{root_rockk_setup}
\end{figure}

\begin{figure}[tb]
	\centering	
%
%

\definecolor{mycolor3}{rgb}{0.50196, 0.50196, 0.00000}
\begin{tikzpicture}[scale=0.6]

\begin{axis}[%
width=4.521in,
height=3.566in,
at={(0.758in,0.481in)},
scale only axis,
xmin=0.5,
xmax=4.5,
xtick={1,2,3,4},
xticklabel style={font=\fontsize{16}{16}\selectfont, color=white!15!black},
xticklabels={{\shortstack{BAA\\(Pebbles)}},{\shortstack{BAA\\(root)}},{\shortstack{BPA\\(root)}},{\shortstack{BPA\\(Pebbles)}}},
ymin=0,
ymax=100,
yticklabel style={font=\fontsize{16}{16}\selectfont, color=white!15!black},
ylabel style={font=\fontsize{18}{18}\selectfont, color=white!15!black},
ylabel={SM [\%]},
axis background/.style={fill=white}
]
\addplot [color=black, dashed, forget plot]
  table[row sep=crcr]{%
1	36.8524658551594\\
1	45.5797381434851\\
};
\addplot [color=black, dashed, forget plot]
  table[row sep=crcr]{%
2	28.861990799051\\
2	34.5675750173053\\
};
\addplot [color=black, dashed, forget plot]
  table[row sep=crcr]{%
3	40.7769849717827\\
3	44.8602254511118\\
};
\addplot [color=black, dashed, forget plot]
  table[row sep=crcr]{%
4	42.9704129982709\\
4	47.6114047088062\\
};
\addplot [color=black, dashed, forget plot]
  table[row sep=crcr]{%
1	25.3198464865926\\
1	30.733050993215\\
};
\addplot [color=black, dashed, forget plot]
  table[row sep=crcr]{%
2	20.6653726481906\\
2	25.046210878464\\
};
\addplot [color=black, dashed, forget plot]
  table[row sep=crcr]{%
3	34.2257228594141\\
3	38.0431033656123\\
};
\addplot [color=black, dashed, forget plot]
  table[row sep=crcr]{%
4	36.0347777345785\\
4	39.8702144461219\\
};
\addplot [color=black, forget plot]
  table[row sep=crcr]{%
0.875	45.5797381434851\\
1.125	45.5797381434851\\
};
\addplot [color=black, forget plot]
  table[row sep=crcr]{%
1.875	34.5675750173053\\
2.125	34.5675750173053\\
};
\addplot [color=black, forget plot]
  table[row sep=crcr]{%
2.875	44.8602254511118\\
3.125	44.8602254511118\\
};
\addplot [color=black, forget plot]
  table[row sep=crcr]{%
3.875	47.6114047088062\\
4.125	47.6114047088062\\
};
\addplot [color=black, forget plot]
  table[row sep=crcr]{%
0.875	25.3198464865926\\
1.125	25.3198464865926\\
};
\addplot [color=black, forget plot]
  table[row sep=crcr]{%
1.875	20.6653726481906\\
2.125	20.6653726481906\\
};
\addplot [color=black, forget plot]
  table[row sep=crcr]{%
2.875	34.2257228594141\\
3.125	34.2257228594141\\
};
\addplot [color=black, forget plot]
  table[row sep=crcr]{%
3.875	36.0347777345785\\
4.125	36.0347777345785\\
};
\addplot [color=black, forget plot]
  table[row sep=crcr]{%
0.75	30.733050993215\\
0.75	36.8524658551594\\
1.25	36.8524658551594\\
1.25	30.733050993215\\
0.75	30.733050993215\\
};
\addplot [color=black, forget plot]
  table[row sep=crcr]{%
1.75	25.046210878464\\
1.75	28.861990799051\\
2.25	28.861990799051\\
2.25	25.046210878464\\
1.75	25.046210878464\\
};
\addplot [color=black, forget plot]
  table[row sep=crcr]{%
2.75	38.0431033656123\\
2.75	40.7769849717827\\
3.25	40.7769849717827\\
3.25	38.0431033656123\\
2.75	38.0431033656123\\
};
\addplot [color=black, forget plot]
  table[row sep=crcr]{%
3.75	39.8702144461219\\
3.75	42.9704129982709\\
4.25	42.9704129982709\\
4.25	39.8702144461219\\
3.75	39.8702144461219\\
};
\addplot [color=red, forget plot]
  table[row sep=crcr]{%
0.75	33.3625657126903\\
1.25	33.3625657126903\\
};
\addplot [color=red, forget plot]
  table[row sep=crcr]{%
1.75	26.7386948596\\
2.25	26.7386948596\\
};
\addplot [color=red, forget plot]
  table[row sep=crcr]{%
2.75	39.2898894155911\\
3.25	39.2898894155911\\
};
\addplot [color=red, forget plot]
  table[row sep=crcr]{%
3.75	41.3841411757287\\
4.25	41.3841411757287\\
};
\addplot [color=black, only marks, mark=+, mark options={solid, draw=red}, forget plot]
  table[row sep=crcr]{%
1	46.0705232308262\\
1	46.1667425870812\\
1	46.2048307880566\\
1	46.2302937161276\\
1	46.2616359752888\\
1	46.3986881375969\\
1	46.6718894013591\\
1	46.7768037833152\\
1	46.8006560406345\\
1	46.8104462461158\\
1	47.0435354857288\\
1	47.3092701234856\\
1	47.8959508400311\\
1	47.9277497396295\\
1	48.5890205395272\\
1	48.7352958492817\\
1	48.835781157349\\
1	48.8478257562418\\
1	49.0006974413279\\
1	49.0847583340173\\
1	49.3585655450676\\
1	49.5414600467034\\
1	50.0782539785552\\
1	50.4753648078873\\
1	50.685056995472\\
1	50.7258962435166\\
1	51.0931372328007\\
1	51.4402749814877\\
1	51.4938225281001\\
1	51.6690499213673\\
1	51.7003564048931\\
1	52.1616248940804\\
1	52.5895455526471\\
1	53.0728160321745\\
1	53.2257597245623\\
1	53.3742369861268\\
1	53.5785060779439\\
1	53.5971268589205\\
1	53.800113107202\\
1	54.0430602794494\\
1	56.0013689945627\\
1	62.1068304109055\\
1	65.8481574323241\\
1	74.8979571386165\\
};
\addplot [color=black, only marks, mark=+, mark options={solid, draw=red}, forget plot]
  table[row sep=crcr]{%
2	34.807209840497\\
2	35.1236567829734\\
2	35.1360651778715\\
2	35.1796346167711\\
2	35.3911168375759\\
2	35.5361924880861\\
2	35.7430314990188\\
2	35.8111914824585\\
2	35.8349010179934\\
2	35.8368961445164\\
2	35.9801640068081\\
2	36.0956008347777\\
2	36.3908258039736\\
2	36.4708981327471\\
2	37.3134455568079\\
2	37.6081804020767\\
};
\addplot [color=black, only marks, mark=+, mark options={solid, draw=red}, forget plot]
  table[row sep=crcr]{%
3	33.1837846211791\\
3	44.9745845324422\\
3	45.1753615896448\\
3	45.3169742164029\\
3	45.3572719883693\\
3	45.4412301523033\\
};
\addplot [color=black, only marks, mark=+, mark options={solid, draw=red}, forget plot]
  table[row sep=crcr]{%
4	47.9122349127172\\
4	49.0185377161744\\
4	49.6125774491289\\
};
\addplot [color=mycolor3, dashed, line width=2.0pt, forget plot]
  table[row sep=crcr]{%
0.5	62.5\\
4.5	62.5\\
};
\end{axis}

\begin{axis}[%
width=5.833in,
height=4.375in,
at={(0in,0in)},
scale only axis,
xmin=0,
xmax=1,
ymin=0,
ymax=1,
axis line style={draw=none},
ticks=none,
axis x line*=bottom,
axis y line*=left
]
\end{axis}
\end{tikzpicture}%
	\caption{\small SM estimation in the presence of plant roots and pebbles as the soil clutters. }
	\label{root_rock_resulat}
\end{figure}
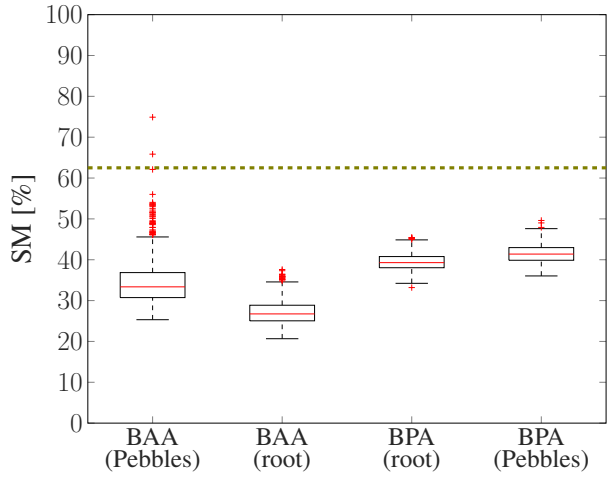

In order to create a more realistic scenario, we updated the test bed to incorporate the impact of roots and pebbles, as depicted in Fig. \ref{root_rockk_setup}. 
In this scenario, we consider a soil bag of $60\%$ SM and utilize the developed model with the statistical parameters obtained from exponential fitting,  detailed in Tab. \ref{ab charactiriz}.
The results of SM estimation for $1000$ model generations are illustrated in Fig. \ref{root_rock_resulat}. 
In this analysis, we applied the SVD approach for clutter reduction by discarding $2$ SVs, rather than $1$ SV, to mitigate the impact of added clutter.
Remarkably, this demonstrates SMEEs quite similar to the clutter-free medium for both imaging algorithms.

\subsection{Laboratory SDI results}
\begin{figure}[tb]
	
	\subfloat[]{\input{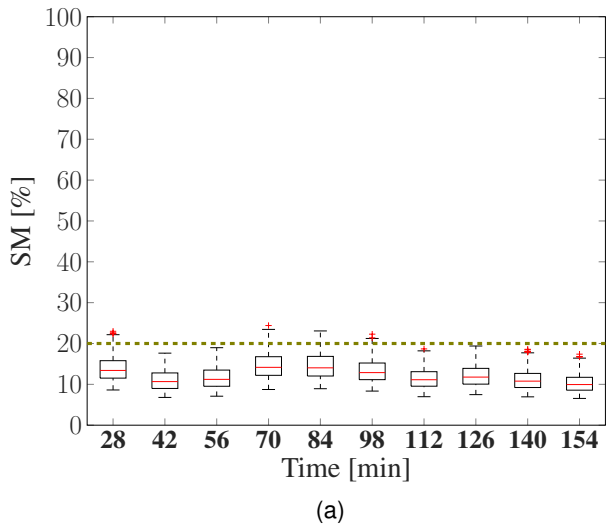}}
	
	\subfloat[]{\input{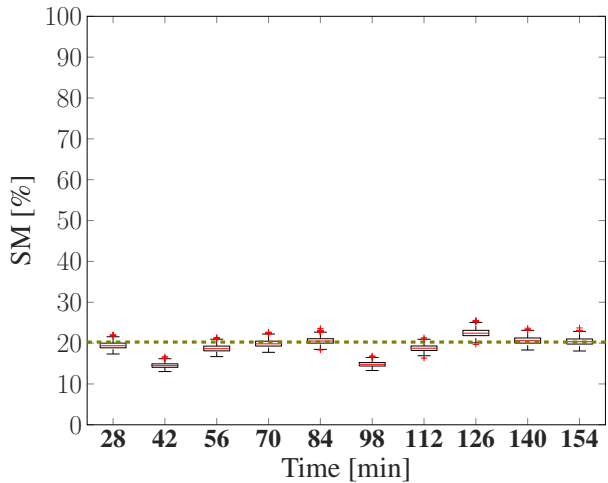}}
	
	\caption{Temporal SM estimation for a setup mimicking an SDI system depicted for (a) BAA and (b) BPA. Dashed line shows the measured exact value of SM at the last time point using the gravimetric method.}
	\label{real_leak_result}
\end{figure}

Fig. \ref{real_leak_result} illustrate the  SM for a leaky pipe as a case in SDI systems. 
We conduct $10$ scans of the test box over a time duration of \SI{154}{min}, and apply the proposed approach to estimate the SM content through $1000$ generations of proposed statistical model for each scan.
To get the real value of SM average for the verification of estimated SMs, we segregated a sample of soil from the moistened area and used the gravimetric method. This method involves drying the soil in an oven and measuring the weight difference before and after drying enabling us to obtain the weight of water in the soil and  SM content.
As depicted, after $1$ hour, the SM reaches the saturated value of $20\%$, which closely matches the real measured value particularly for BPA.

\section{Conclusion }\label{conclusion}
Our research highlights the promising utilization of microwave imaging systems (MIS) within subsurface drip irrigation (SDI) for agricultural applications. We employ a model tailored to various moisture levels, derived from the subspace information of moist area images, to estimate moisture content accurately. Our methodology utilizes back projection (BPA) and Born approximation (BAA) algorithms for image formation, leveraging a wide frequency range for data acquisition and statistical curve fitting to construct the model. Through experimentation in a laboratory-scaled SDI system, we achieved real-time soil wetness monitoring.
Significantly, our results demonstrate that BPA surpasses BAA, yielding a lower soil Soil Moisture(SM) estimation error. Additionally, the developed model exhibits robustness against medium clutter such as plant roots and pebbles.
Looking ahead, there is potential to enhance the current setup into portable equipment with faster data capturing considering frequency modulation continuous waves. Furthermore, integrating the processing tools introduced in our study with airborne radar systems could improve scanning efficiency for larger areas.
The successful integration of this technology into agricultural practices holds promise for optimizing water usage and fostering sustainable and productive crop cultivation.

\bibliographystyle{IEEEtran}
\bibliography{IEEE-TAP-latex-template}

\vfill

\end{document}